\documentclass[aps,twocolumn,pra,superscript,floatfix,superscriptaddress,showpacs]{revtex4-1}
\usepackage{amssymb}

\usepackage{epsf}
\usepackage{epsfig}
\usepackage{graphicx}
\usepackage{dcolumn}
\usepackage{braket}
\usepackage{bm}
\usepackage{amsfonts}
\usepackage{amsmath}
\usepackage{amssymb}
\usepackage{color,soul}
\usepackage{wasysym}
\usepackage{mathrsfs}
\usepackage{natbib}
 \usepackage{multirow}
\usepackage{times}

\newcommand{\bea}{\begin{eqnarray}}
\newcommand{\eea}{\end{eqnarray}}

\newcommand{\ci}{\mathrm{i}}

\begin{document}

\title{Floquet topological transitions in a driven one-dimensional topological insulator}

\author{V. Dal Lago}
\affiliation{Instituto de F\'{\i}sica Enrique Gaviola (CONICET) and FaMAF, Universidad Nacional de C\'ordoba, Argentina}
\author{M. Atala}
\affiliation{Laboratorium f\"ur Physikalische Chemie, ETH Z\"urich, 8093 Z\"urich, Switzerland}
\author{L. E. F. Foa Torres}
\affiliation{Instituto de F\'{\i}sica Enrique Gaviola (CONICET) and FaMAF, Universidad Nacional de C\'ordoba, Argentina}

\begin{abstract}
The Su-Schrieffer-Heeger model of polyacetylene is a paradigmatic Hamiltonian exhibiting non-trivial edge states. By using Floquet theory we study how the spectrum of this one-dimensional topological insulator is affected by a time-dependent potential. In particular, we evidence the competition among different photon-assisted processes and the native topology of the unperturbed Hamiltonian to settle the resulting topology at different driving frequencies. While some regions of the quasienergy spectrum develop new gaps hosting Floquet edge states, the native gap can be dramatically reduced and the original edge states may be destroyed or replaced by new Floquet edge states. Our study is complemented by an analysis of Zak phase applied to the Floquet bands. Besides serving as a simple example for understanding the physics of driven topological phases, our results could find a promising test-ground in cold matter experiments.
\vspace{0.5cm}
\end{abstract}

\date{\today}
\maketitle

\section{Introduction}

Since the discovery of the integer quantum Hall effect \cite{vonKlitzing1980}, the physics of topological states has established itself as a privileged crossroads for diverse communities. A quarter of a century later, the discovery of topological insulators \cite{Koenig2007} has propelled the interest in this area \cite{Moore2010,Hasan2010}. This time, rather than  high magnetic fields, the key ingredient was the spin-orbit coupling which is naturally strong in compounds made of heavy elements. Recent studies hint that there might be a third alternative enabling non-trivial topology: shining a laser on a sample \cite{Oka2009,Kitagawa2011,Lindner2011,Perez-Piskunow2014} to generate \textit{Floquet topological states} \cite{Lindner2011,Cayssol2013,Bukov2014,TenenbaumKatan2013a,TenenbaumKatan2013}. The sample could be either graphene \cite{Oka2009,Kitagawa2011,SuarezMorell2012,Perez-Piskunow2014}, a trivial insulator \cite{Lindner2011}, 3D Dirac semimetal \cite{Narayan2015}, or a topological insulator \cite{Calvo2015}. The opening of laser-induced bandgaps has already been observed \cite{Wang2013} and new experiments \cite{Rechtsman2013,Jotzu2014,Gao2015} and theories are being developed to unveil the states' fingerprint of their topology \cite{Perez-Piskunow2014,Usaj2014,Dahlhaus2014,Thakurathi2013,Sentef2015}, \textit{e.g.} the bulk boundary correspondence \cite{Rudner2013,Ho2014,Carpentier2015}, their dynamics \cite{Goldman2014,DAlessio2014}, occupation \cite{Seetharam2015,Iadecola2015}, transport properties \cite{Kundu2014,Dehghani2014,Farrell2015,Dehghani2015} and the connection between the Hall response and the edge states \cite{Foa2014}.

\begin{figure}[tbph]
\includegraphics[width=0.85\columnwidth]{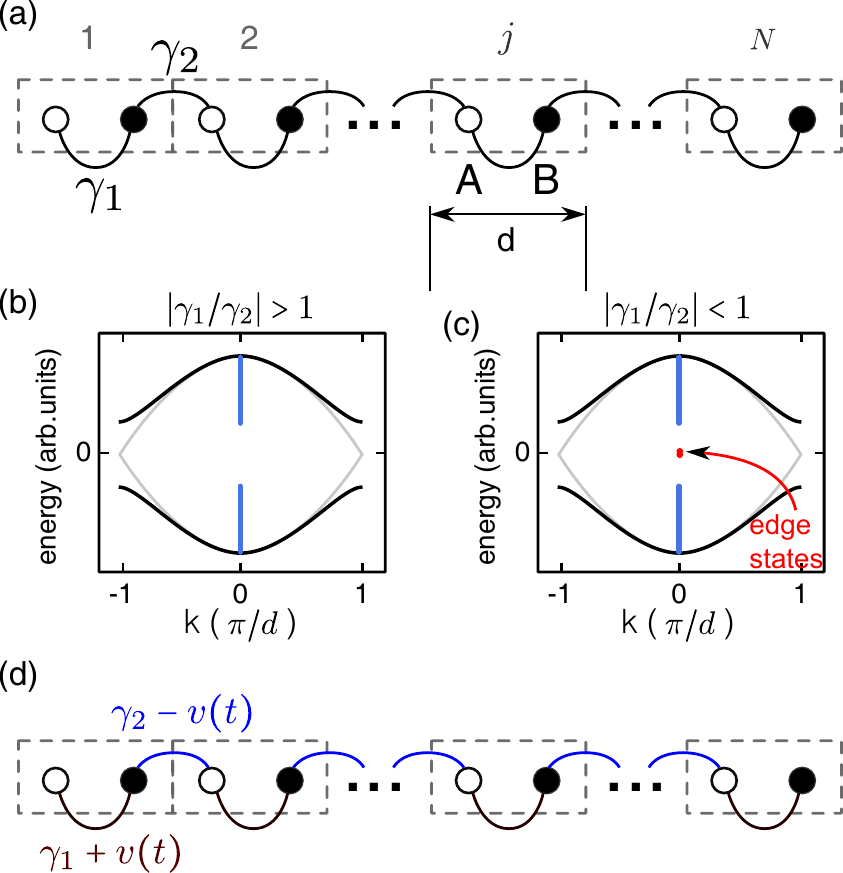}
\caption{(color online) (a) Scheme representing a finite section of the Su-Schrieffer-Heeger model with $N$ unit cells and where $\gamma_1$ and $\gamma_2$ are the intra- and intercell hoppings, respectively. While the bulk spectrum is independent of the ratio $|\gamma_1/\gamma_2|$ (shown in (b) and (c) in gray for $\delta=0$ and black line for $\delta \ne 0$), the spectrum of a finite chain (shown in blue dots in panels (b) and (c)) differs in the absence (b,  $|\gamma_1/\gamma_2|>1$) or presence (c,  $|\gamma_1/\gamma_2|<1$) of mid-gap states. (d) Scheme of the driven SSH model considered in the text.}
\label{fig1}
\end{figure}

At the time of the birth of the integer quantum Hall effect \cite{vonKlitzing1980,Thouless1982} another important finding was the discovery of conducting polymers \cite{Su1979,Heeger1988}.
Later on, it became clear that the simple tight-binding model proposed by Su, Schrieffer and Heeger \cite{Su1979} in 1979 to describe the dimerization in polyacetylene (today named the SSH model) is, indeed, a minimal example of a one-dimensional topological insulator (see for example Refs. \onlinecite{Fefferman2014,Li2014}). The topologically trivial or non-trivial character of the dimerized chain is controlled by the relative strength of the intracell-to-intercell couplings: A tight-binding chain with hoppings bearing alternating values $\gamma_1$ and $\gamma_2$ ($\gamma_1$ being the intracell hopping as represented in Fig. \ref{fig1}(a)) sustains one topological state localized at each termination if $|\gamma_1/\gamma_2|<1$, and zero otherwise (as shown in Fig. \ref{fig1} (c) and (b)). The existence of such states, in turn, is related to a topological invariant called Zak phase.

Here we inquire about the effect of a time-dependent perturbation on the SSH model, thereby providing a minimal case of a driven topological insulator. The driving considered here consists of a modulation of the hoppings with spacial period $d$ (the same as for the SSH model, see scheme in Fig. \ref{fig1}(d)). In particular, a relevant question is how driving affects the native topology. For example, if starting from the undimerized chain one adds a time-dependent modulation of the hoppings, then, as the instantaneous ratio $|\gamma_1/\gamma_2|$ changes, snapshots taken at different times would correspond to configurations consistent with a switching from a topologically trivial insulator to a non-trivial one, back and forth. Although previous works have addressed driven one-dimensional systems in a variety of contexts \cite{Asboth2014,Seetharam2015}, the above question still remain.

By using Floquet theory we show that driving can induce new gaps depending critically on the driving frequency/strength and change dramatically the gaps of the undriven model. Interestingly, the newly formed gaps may host Floquet edge states while those at the pre-existent gaps can be annihilated by the driving. The topological transitions leading to such changes are studied in detail complementing our numerical study with a calculation of the relevant topological invariant, i.e. the Zak phase \cite{Zak1989}. Our results reveal a subtle competition between the different available photon-assisted processes and the native topology of the model to set the final character of each gap. We show that as the frequency is lowered, each time new inelastic processes come to play one has that either that the pre-existent edge states are destroyed, if they already exist, or new ones are created, if they  were not present before. These Floquet topological transitions could be tested in cold matter where there is much interest in driven topological phases from the theoretical point of view \cite{Choudhury2014,Goldman2015,Quelle2015} and where recent experimental progress has allowed the realization of the SSH model \cite{Atala2013}, the Haldane model \cite{Jotzu2014}, the Hofstadter model\cite{Aidelsburger2015}, and even the measurement of the relevant topological invariants \cite{Atala2013,Aidelsburger2015}.

\section{Floquet approach and opening of bandgaps in the quasienergy spectrum}
\subsection{Floquet theory applied to the driven SSH model} 
The Su-Schrieffer-Heeger (SSH) model \cite{Su1979} describes the dimerization that occurs in a one-dimensional periodic system subject to a cell doubling perturbation in the spirit of the Peierls transition\cite{Peierls1955,PeierlsMoreSurprises}. Here we consider a variant of the SSH model where the hoppings are modulated in time so that nearby bonds have opposite phases:

\begin{eqnarray}
\label{Hamiltonian}
& &{\cal H}= - \sum_{j} \big( \gamma^{ }_1 c^{\dagger}_{A,j} c^{ }_{B,j}  + \gamma^{ }_2 c^{\dagger}_{B,j-1} c^{ }_{A,j} +h.c. \big) + \nonumber \\
& & + \sum_{j} \big(v(t) c^{\dagger}_{A,j} c^{ }_{B,j}  - v(t) c^{\dagger}_{B,j-1} c^{ }_{A,j} +h.c. \big), \nonumber
\end{eqnarray}
where $c^{\dagger}_{\alpha,j}$ and $c^{}_{\alpha,j}$ are the creation (anihilation) operators at site $\alpha$ (which can be either $A$ or $B$-type) of the $j$-th unit cell, $\gamma_1= \gamma_0+\delta$ and $\gamma_2= \gamma_0-\delta$, $\delta$ being the dimerization strength and $v(t)=2 V_{{\rm ac}} \cos(\Omega t)$ with $V_{\rm{ac}}$ being the driving amplitude. The onsite energies $E^{ }_{\alpha,j}$ may all be taken equal to a reference energy (zero energy), $\gamma_0$ is taken as the unit of energy and the lattice constant for the dimerized phase is $d$ (as represented in Fig. \ref{fig1}(a)). 

The first term on the rhs of Eq. (\ref{Hamiltonian}) corresponds to the usual SSH model and the second term accounts for the driving. In the bulk limit, the static part of the Hamiltonian can also be written as ${\cal H}_{k}=\bf{h_{k}}.\sigma$, where ${\cal H}_{k}$ is a $2\times2$ matrix written in the basis of A and B sites, $\sigma$ is the vector of Pauli matrices $(\sigma_x,\sigma_y,\sigma_z)$ and $\bf{h_{k}}$ is a vector with vanishing z-component since A and B have the same local energy. The latter assures that chiral symmetry is preserved \cite{Franz2013}: if $+\varepsilon$ is an eigenvalue then its opposite $-\varepsilon$ is one too (the spectrum is therefore symmetric). Imposing this symmetry ensures the existence of two distinct topological phases ($\delta>0$ and $\delta<0$) for the static system. One can verify that the driving term satisfies a generalized chiral symmetry and the Floquet spectrum is symmetric with respect to $\varepsilon=m\hbar\Omega/2$ ($m$ integer).

The properties of interest in our work can be obtained by using Floquet theory which is particularly tailored for time-periodic Hamiltonians \cite{Grifoni1998,Kohler2005}. Given a time-periodic Hamiltonian with period $T$, there is a complete set of Floquet solutions of the form $\psi_{\alpha}(\bm{r},t)=\exp(-\ci\varepsilon_{\alpha}t/\hbar) \phi_{\alpha}(\bm{r},t)$, where ${\varepsilon_{\alpha}}$ are the so-called quasienergies and $\phi_{\alpha}(\bm{r},t+T)=\phi_{\alpha}(\bm{r},t)$ are the associated Floquet states. By inserting these solutions into the time-dependent Schr\"odinger equation, one gets that the Floquet states satisfy an equation analogous to the time-independent Schr\"odinger equation with the Hamiltonian being replaced by the Floquet Hamiltonian $\hat{\cal{H}}_F\equiv\hat{\cal{H}}-\ci\hbar \frac{\partial}{\partial t}$. Thus, one has an eigenvalue problem in the direct product Floquet space \cite{Sambe1973}: $\mathscr{R}\otimes \mathscr{T}$, $\mathscr{R}$ being the usual Hilbert space and $\mathscr{T}$ the space of periodic functions with period $T=2\pi/\Omega$ which is spanned by the functions $\exp(i n\Omega t)$. The index $n$ can be assimilated to the number of `photon' excitations \cite{Shirley1965}, and defines a subspace also called \textit{$n^{th}$ Floquet replica}. Note that in this energy domain solution we do not need to resort to the consideration of the time-evolution operator.

In our calculations we used different codes including: a code built on Kwant \cite{Groth2014} and a home-made implementation using recursive Green's functions allowing to deal with a semi-infinite system. The numerical calculation of the Zak phase was carried out using the Python Tight-Binding package (PythTB).

\subsection{Driving-induced gaps in the bulk spectrum} 

\begin{figure}[tbph]
\centering
\includegraphics[width=\columnwidth, angle=0]{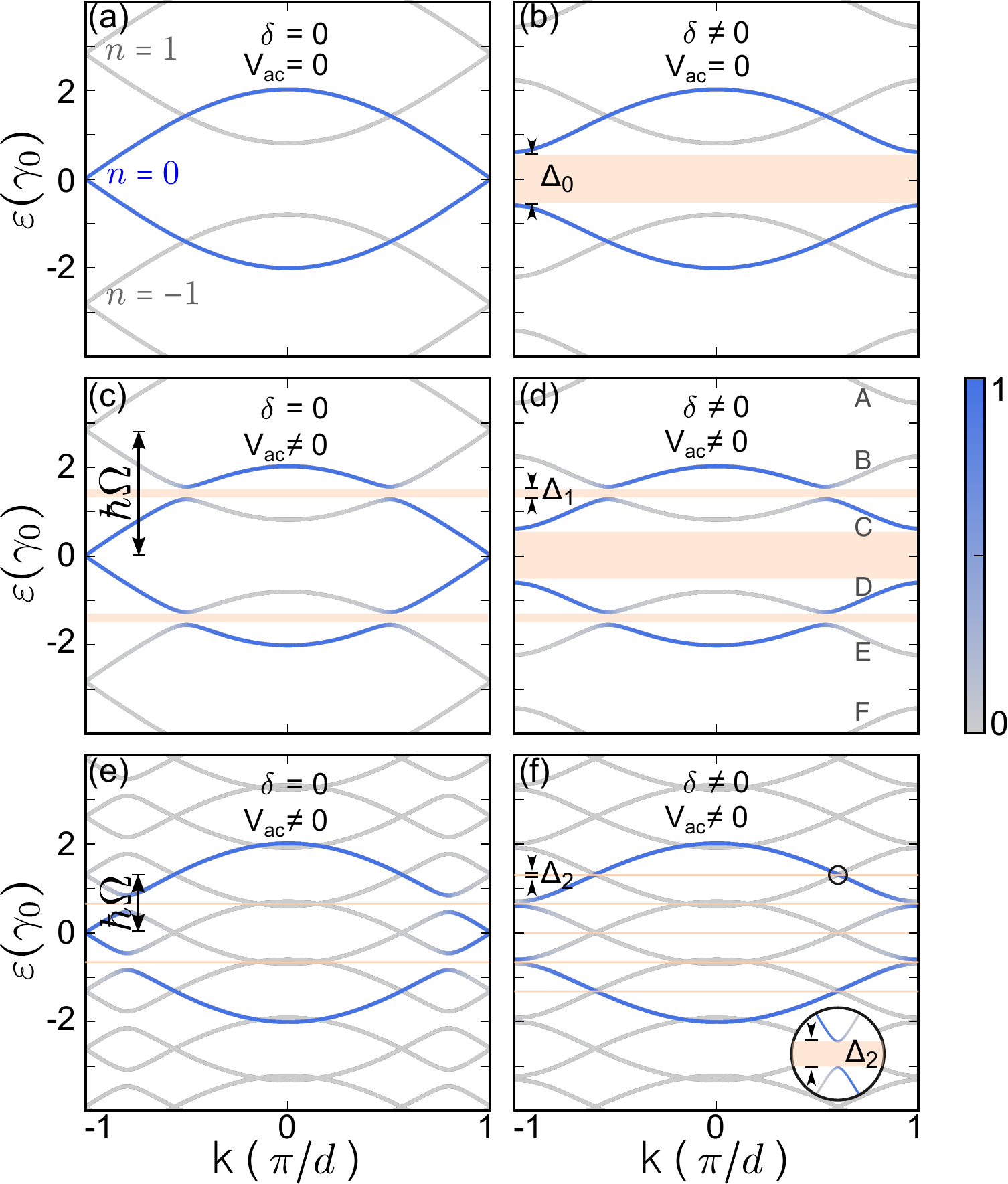}
\caption{(color online) Quasienergy spectra for the (bulk) Su-Schrieffer-Heeger model obtained for different set of values of the dimerization constant $\delta$ and the driving amplitude $V_{\rm{ac}}$. In the available range we can distinguish the Floquet replicas with $n=-1,0,1$, and the color bar indicates the contribution to the time-averaged density of states. Starting from the undriven model (a) and (b), and considering the process related to the $n=0$ replica, we can see how the model develops gaps at $\pm\hbar\Omega/2$ in (c) and (d), and also at $\pm\hbar\Omega$ in (f). In these plots $\hbar\Omega=2.8\gamma_0$ for (c) and (d) and $\hbar\Omega=1.3\gamma_0$ for (e) and (f); $\delta=0.3\gamma_0$ in (b),(d) and (f); and $V_{\rm{ac}}=0.1\gamma_0$ (c) to (f).}
\label{fig2}
\end{figure}

Let us focus first on the quasienergy spectrum of the bulk system. Figures \ref{fig2} (a) and (b) show the spectrum of the undriven system without ($\delta=0$) and with ($\delta \neq 0$) the cell-doubling perturbation, respectively. While the former does not have a gap, the latter does have one at zero energy of magnitude $\Delta_0=4|\delta|$. The color scale represents the weight on different replicas, from blue for unit weight on the $n=0$ replica to grey. This weight can be interpreted as the contribution of each state of energy $\varepsilon$ to the time-averaged density of states at the same energy \cite{Oka2009}.

The most relevant effects of driving take place at the points where the spectrum becomes degenerate. For simplicity, we will focus only on the energy range spanned by the $n=0$ replica. In that case, the degenerate points at the frequency plotted in Fig. \ref{fig2} (a-d) are located at either $\varepsilon=0$ or $\varepsilon\pm \hbar\Omega/2$.
 
Once the time-dependent perturbation is switched on (as in Figs. \ref{fig2} (c) and (d)), the degeneracies at $\pm \hbar\Omega/2$ (involving states with $n=0$ and $n=\pm1$) are lifted leading to the \textit{driving-induced bandgaps}. Similar gaps were previously found in the context of carbon nanotubes affected by the interaction with a single optical phonon mode \cite{FoaTorres2006,FoaTorres2008}, which gives a quantized version of our model.

What is the origin of such gaps in this one-dimensional system? In contrast to the driving-induced gaps in graphene illuminated with a circularly polarized laser \cite{Oka2009,Calvo2011}, where opening a gap among the $n=0$ and $n=\pm1$ replicas requires breaking time-reversal symmetry (TRS), \textit{the gap opening mechanism in the driven SSH model relies on the spacial periodicity of the time-dependent perturbation} (no TRS breaking is necessary). Indeed, the time-dependent perturbation must have a component with a spacial period $d$ to mix different Floquet replicas, so that the degeneracies between states at $\pm\hbar\Omega/2$ are lifted as shown in Figs. \ref{fig2}(c) and (d). The gap has a magnitude of $\Delta_1\simeq 4|V_{\rm{ac}}|$. Since this process involves the Floquet replicas with $n=0$ and $n=\pm 1$, energy exchange (photon emission/absorption) is crucial for these gaps to occur, thereby highlighting their \textit{non-adiabatic} nature.

When the frequency is low enough so that more replicas overlap with each other, one can notice how higher order effects develop (new degeneracies appear within the spectral support of the $n=0$ replica). This is the case in Figs. \ref{fig2}(e) and (f) where $\hbar\Omega=1.1\gamma_0$. The gaps at $\pm \hbar\Omega/2$ are almost unnoticeable in this case and there are new driving induced gaps at $\varepsilon=\pm\hbar\Omega$ which appear only when $\delta\neq0$ (see panels (e) and (f)) and correspond to the mixing of the $n=0$ and $n=\pm 2$ replicas. The processes leading to those gaps are more subtle and involve virtual transitions through intermediate states. The magnitude of the gap, that we denote with $\Delta_2$, can be worked out analytically. One gets: $\Delta_2\simeq(\Delta_1/(2\hbar\Omega))^2\Delta_0$, hence, the gap is proportional to $V_{\rm{ac}}^2\times \Delta_0$. The gap at zero energy (which is reduced as compared to panels (c) and (d)) is also determined now by $\Delta_2$.

\section{Floquet edge states, Topological Transitions and Zak phase}

\begin{figure*}[t]
\centering
\includegraphics[width=\textwidth]{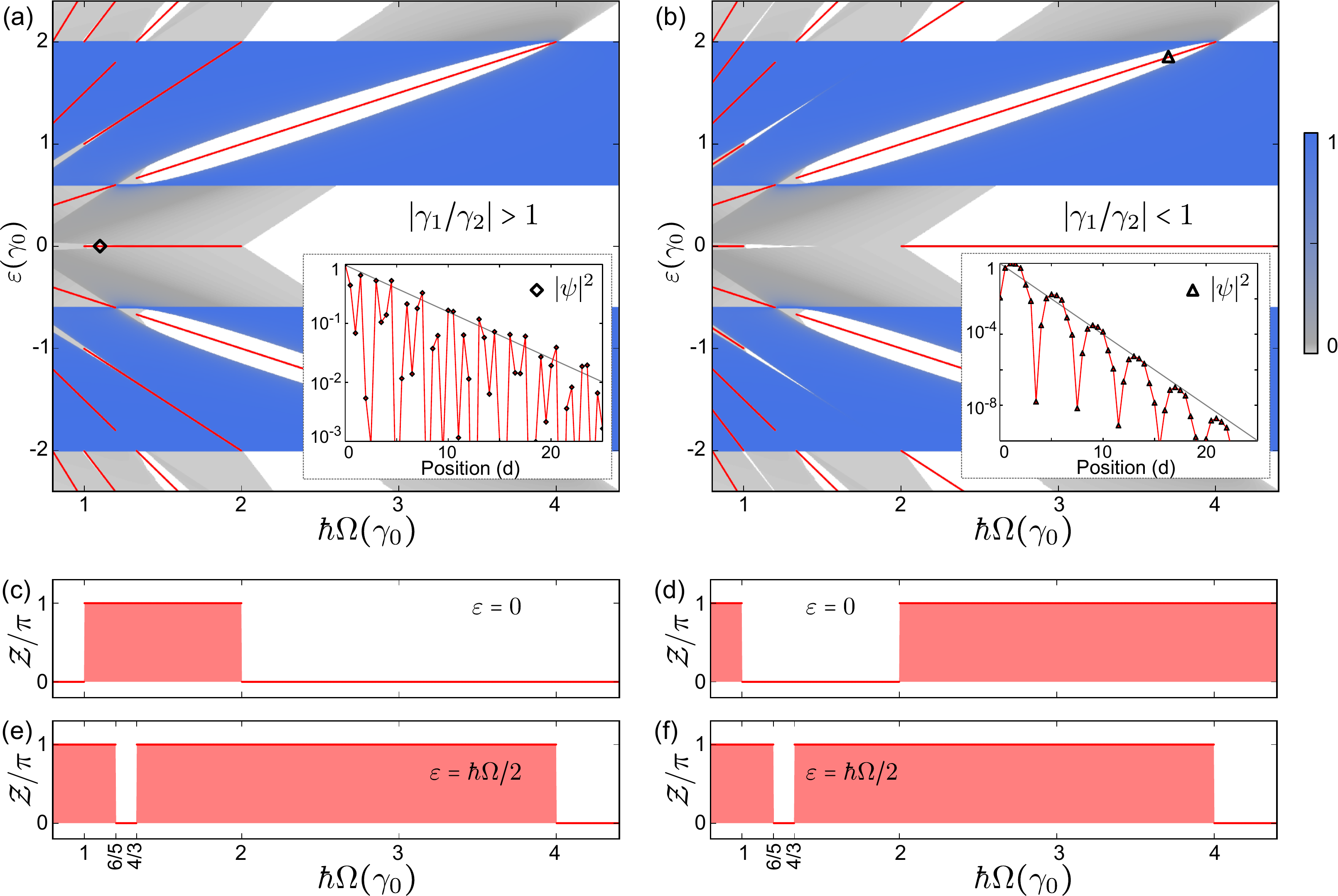}
\caption{(color online) Floquet spectrum ((a) and (b)) and Zak phases (c-f) as a function of the driving frequency $\Omega$ from 0.8 to 4.4 $\gamma_0$. The calculations correspond to $V_{\rm{ac}}=0.1\gamma_0$ and $\delta=+0.3\gamma_0$ (panels (a), (c) and (e)) and $\delta=-0.3\gamma_0$ (panels (b), (d) and (f)). The color scale from grey to blue in panels (a) and (b) indicates the weigth of the corresponding bulk state on the $n=0$ Floquet subspace (white indicates the absence of states, i.e. the existence of a gap in the bulk spectrum). The midgap states are all shown in red, irrespective of their weight on a particular Floquet replica. These edge states are localized at the end of the system. The modulus squared of the the eigenvector components as a function of position for two of such edge states marked with a diamond in (a) and up triangle in (b) are shown in the corresponding insets (scatters). (a)-inset shows the modulus squared on the $n=+1$ replica, which is the same as the one on $n=-1$, while (b)-inset shows the weight on the $n=0$ replica (which is numerically equal to that on the $n=+1$ replica). They follow an exponential decay (grey lines) with a decay length inversely proportional to the corresponding gap. (c) and (d) ((e) and (f)) show the Zak phase for the sates filled up to $\varepsilon=0$ ($\varepsilon=\hbar\Omega/2$). Up to 13 replicas are used for the calculations in the range shown in the figures.}
\label{fig3}
\end{figure*}

The analysis of the previous section gives a first hint on the spectrum of the driven SSH model but does not reveal its most interesting face: Are the Floquet gaps topological? Do we have Floquet topological edge states? How does this depend on the topology of the undriven model?
To answer these questions we analyze the Floquet spectrum of a semi-infinite system in a wide range of frequencies  chosen to match the experimentally relevant regime in cold matter experiments. Due to the existence of an edge one can infer on the topology from the eventual presence of edge states (a result which we will later on confirm based on the Zak phase). The results are shown in Figs. \ref{fig3}(a) for $|\gamma_1/\gamma_2|>1$ and \ref{fig3}(b) for $|\gamma_1/\gamma_2|<1$. At each frequency, the Floquet spectrum is plotted on the vertical scale. For easier interpretation 
a (blue) color scale corresponding to the weight of the states on the $n=0$ Floquet subspace is used (thus, it gives the time-averaged density of states for that electronic energy). On the other hand, edge states are plotted in red, irrespective of their weight on a particular replica, this way one can spot them in the full range without zooming-in the plot. By following the midgap states with the driving frequency, Figure \ref{fig3} resembles the \textit{fan diagrams} widely used for Landau levels as a function of the magnetic field.

In the following we discuss first the set of relevant frequencies where the topology may change (the \textit{transition points}). We then analyze these transitions by looking at the presence of edge states and Zak phase. As we will see below, whenever a new pair of replicas starts to contribute at a given midgap energy, the Zak phase acquires a $\pi$ shift, therefore destroying the pre-existent edge states or creating new ones.

\subsection{High and low-frequency regimes and transition points}

Figure \ref{fig3} shows how the gaps at $\varepsilon=0,\pm \hbar\Omega/2$ evolve with frequency. It also reveals two very distinct regimes: high and low frequency. For simplicity let us imagine that the driving is turned off while keeping the Floquet picture. Then one has the original energy bands and the replicas, which are displaced by $n\hbar\Omega$, as in Fig. \ref{fig2}(b). The high frequency regime occurs when different Floquet replicas do not overlap at the energy of interest. This is satisfied if the frequency is so large that the replicas are well separated. If we are interested in what happens, say, close to $\varepsilon=0$, as long as $\hbar\Omega>2|\gamma_0|$ one is in the high frequency regime. On the other hand, close to $\varepsilon=\pm\hbar\Omega/2$, the high frequency regime takes place for $\hbar\Omega>4|\gamma_0|$ and the gaps at $\pm \hbar\Omega/2$ do not form.
As the frequency is lowered, the replicas overlap enough with each other so that two or more of them have states close to the energy we are interested in and we get into the low-frequency regime. Whenever a new set of replicas enter into the game (or leave it) defines a potential \textit{transition point}, where the topology of the bandstructure may change.

When lowering the frequency, whenever new bands acquire spectral weight at $\varepsilon=0$ they do it in pairs (because of the electron-hole symmetry of the Hamiltonian). The first of such events occurs at $\hbar\Omega=2|\gamma_0|$ and subsequent ones follow the rule $\hbar\Omega=2|\gamma_0|/m$, $m=1,2,...$. Analogously, the pairs of replicas cease to have spectral weight at frequencies given by $\hbar\Omega=2|\delta|/m$, $m=1,2,...$. A similar behavior is found when looking at $\varepsilon=\pm\hbar\Omega/2$, in this case when lowering the frequency new bands enter into the game whenever $\hbar\Omega=4|\gamma_0|/(2m-1)$ and gets out at $\hbar\Omega=4|\delta|/(2m-1)$. For example, in the range shown in Fig. 3, this occurs with the $n=0$ and $n=1$ replicas at frequencies $4|\gamma_0|$ and $4|\delta|$, respectively. These four equations define the transition points.

In the next paragraphs we will aim to describe the topological transitions at those transition points. But before that, let us examine when does the spectrum remains gapped and when it does not. This is, of course, unless all the Floquet replicas present at the energy of interest are simultaneously gapped. This may occur only at precise energies where the global Floquet spectrum is symmetric, in our case  $\varepsilon=m\hbar\Omega/2$ with $m$ an integer.
In particular, at $\varepsilon=0$ in the range $|\gamma_0|<\hbar\Omega<2|\gamma_0|$ (this is, in between the first and the second transition points), the zero energy gap of the $n=0$ replica matches the weaker gap at $\hbar\Omega$ above (below) the center of the $n=-1$ ($n=+1$) replica (with magnitude $\Delta_2$). Thus, the gap is preserved but its width is now set by the smaller one, $\Delta_2$ (as shown in Fig. \ref{fig2} (f)). As the frequency is lowered further and new transition points are traversed, the gap acquires a progressively higher order in $V_{\rm{ac}}$ and therefore decreases to the level where it can be hardly noticed. A similar analysis holds for the gaps at $\pm\hbar\Omega/2$.

\subsection{Topological transitions in the Floquet spectra} 

The topology of the undriven system is controlled by the ratio $|\gamma_1/\gamma_2|$, being trivial when $|\gamma_1/\gamma_2|>1$ and non-trivial when $|\gamma_1/\gamma_2|<1$. Figures \ref{fig3}(a) and (b) show the driven systems in both situations. We have checked that all the midgap states correspond to edge states, which are indeed absent in a bulk calculation. Two of such states are presented in the insets of Fig. \ref{fig3} (a) and (b), showing an exponential decay with an exponent determined by half the corresponding gap width. As expected, in the high frequency regime, the topology of the driven system corresponds to that of the undriven one. This can be inferred from the gaps and the absence/presence of midgap states: Fig. \ref{fig3}(a) does not have midgap states at zero energy while (b) does. In the low-frequency regime, there may also be midgap states appearing at $\pm m\hbar\Omega/2$. For odd $m$, they turn out to be insensitive to the native topology of the undriven model. In contrast, for even $m$ those gaps appear only when $\delta\neq0$ and their topology depends on that of the undriven model.

Let us analyze in more detail the midgap states. As mentioned before, at $\varepsilon=0$ and high frequency ($\hbar\Omega>2|\gamma_0|$), they follow the same prescription as in the undriven SSH model. At $\hbar\Omega=2|\gamma_0|$, the replicas with $n=\pm1$ acquire a finite spectral weight at that energy so that the gap closes and a smaller gap reopens for $\hbar\Omega\rightarrow2|\gamma_0|^{-}$ (the width being this time $\Delta_2$). By examining the midgap states we see a topological change when traversing the transition point at $\hbar\Omega=2|\gamma_0|$. Indeed, the topological trivial phase of the SSH model ($|\gamma_1/\gamma_2|>1$) becomes non trivial and viceversa. Therefore, in the latter case \textit{driving annihilates the topological edge states present in the undriven system} while in the former it creates new edge states localized on the replicas with $n=\pm1$.

At $\hbar\Omega\le|\gamma_0|$, Floquet replicas with $n=\pm 2$ enter into the game reducing the gap width and changing the topology once again. This is, the Floquet edge states are now absent if $|\gamma_1/\gamma_2|>1$ or they re-emerge if $|\gamma_1/\gamma_2|<1$ (one observes one edge state on each edge, two edge states in total). In the latter case, the nature of the new states is different from the native ones as they have a weight which is predominantly on the $n=\pm 2$ replicas.

If we now look at the midgap states at $\varepsilon =\pm \hbar\Omega/2$, one observes that they share the same topology, irrespective of the one of the undriven system (i.e. both Figs. \ref{fig3}(a) and (b) are equivalent in these gaps). Starting from high frequencies, when $\hbar\Omega$ becomes smaller than $4|\gamma_0|$, the gaps at $\pm\hbar\Omega/2$ open up hosting midgap edge states. This is preserved until the next transition point located at $\hbar\Omega=4|\gamma_0|/3$ when topology changes due to the mixing of Floquet replicas with $n=2$ and $n=-1$. The process is reversed at the following transition point located at $\hbar\Omega=4\delta=1.2\gamma_0$ in Figs. \ref{fig3}(a) and (b) (where the replicas with $n=1$ and $n=0$ cease to have spectral weight at $\varepsilon =\pm \hbar\Omega/2$).

In contrast to the topological transitions predicted in the literature for illuminated graphene \cite{Kundu2014,Gomez-Leon2014,Perez-Piskunow2015} (the reader may find detailed maps in Ref.\onlinecite{Perez-Piskunow2015}), here we have a case where the system bears native topological states (which we show that can be destroyed or even be replaced by new ones) and where the number of possible edge states are restricted to a binary value.

How different are the new edge states as compared with those in the undriven model? For the undriven model a simple and elegant argument can describe all the topological phases providing a flavor on the nature of the edge states. Indeed, by considering the fully dimerized limit (when one of the hoppings is zero) one immediately gets all the possible distinct phases: If $\gamma_2=0$ then there are no edge states, whereas if $\gamma_1=0$ one has one edge state at each end of the chain, fully localized on opposite sublattices. In the driven case this simple argument describes the high frequency regime but fails to accomodate for midgap states at zero energy when $\hbar\Omega<2\gamma_0$ or those at $\pm\hbar\Omega/2$. The finite bandwidth is crucial for the new driven phases to occur since they arise because of its interplay with the photon energy (allowing or restricting inelastic processes). As shown in the insets of Fig.\ref{fig3}, the new edge states have a non-vanishing weight on more than one Floquet replica.

\subsection{Zak phase for the driven SSH model} 

To further confirm the topological nature of these states, we resort to the calculation of  topological invariants. The relevant invariant in our case is the Zak phase \cite{Zak1989} ${\cal Z}$ defined as:
\begin{equation}
{\cal Z} = i\oint dk \left\langle u_{k}|\partial_{k}u_{k}\right\rangle
\; ,
\label{zakphase}
\end{equation}
where $|u_{k}\rangle$ are the cell-periodic Bloch states. The Zak phase is essentially the geometric phase acquired after an adiabatic loop in the Brillouin zone.
This phase has been connected to the existence of edge states in graphene \cite{Delplace2011} and, interestingly, it has been measured in cold matter systems simulating the (undriven) SSH model \cite{Atala2013} and also in acoustic systems \cite{Xiao2015}.

Differences between Zak phases for topologically non-trivial systems in one-dimension are quantized in units of $\pi$ even though the value by itself depends on the choice of the unit cell \cite{Atala2013}. The sum of the Zak phases for all the bands with energy below a given gap indicates the existence (with the relevant cumulative phase being $\pi$) or absence (vanishing cumulative phase) of topological mid-gap states.

To compute the Zak phase for the Floquet quasienergy spectrum one needs to truncate the Floquet space. The number of replicas needs to be chosen so that all relevant transitions at the desired energy are kept. For example, for $\hbar\Omega=2.8\gamma_0$, the analytic calculation of the phases ${\cal Z}_{\alpha}$ for each of the bands marked with letters A-F in Fig. \ref{fig2}(d) (considering just those 3 replicas) gives:

\begin{eqnarray}
&&{\cal Z}_{\alpha} =\left\{
                \begin{array}{ll}
                  0 \,(\alpha=A,F)\\
                  \pi \,(\alpha=B,C,D,E)
                \end{array}
              \right.; \, |\gamma_1/\gamma_2|>1,\\
&&{\cal Z}_{\alpha} =\left\{
                \begin{array}{ll}
                  \pi \,(\alpha=A,F)\\
                  0 \,(\alpha=B,C,D,E)
                \end{array}
              \right.;\,|\gamma_1/\gamma_2|<1.
\end{eqnarray}
One verifies that the sum of the Zak phases do correspond with the edge states found in the simulation of Fig.\ref{fig3} (a) and (b). We have also verified this by numerical calculation in the frequency range shown in Fig. \ref{fig3}. This is shown in Figs. \ref{fig3}(c-f), where panels (c) and (d) show the results for the cumulative phase up to $\varepsilon=0$, while (e) and (f) are for $\varepsilon=\hbar\Omega/2$, for the trivial and non-trivial undriven system respectively.

Considering both the results of the Floquet spectrum and of the Zak phase, and assuming that the bulk boundary correspondence holds at all frequencies, the behavior of the driven system both at zero energy and at $\pm\hbar\Omega/2$ can be put in a nutshell with a simple argument: \textit{When lowering the driving frequency, every pair of new replicas entering into the game at a given energy adds a $\pi$ to the cumulative Zak phase, thereby switching the topology from trivial to non-trivial and viceversa.}

\section{Final remarks.}
In summary, we analyze the influence of driving on a one-dimensional topological insulator given by the SSH model. Our analysis reveals the creation of driving-induced bandgaps at $\pm m \hbar\Omega/2$ ($m$ integer) and their evolution. More interestingly, we show that the topology of these bands turns out to exhibit transitions as the driving frequency changes. Both the native gap and the gaps at $\varepsilon = \pm \hbar\Omega/2$ switch topology from non-trivial to trivial and viceversa. We attribute these topological transitions to the competition between the native topology and the one due to the driving and also among different photon-assisted processes themselves. Our numerical results are supported by an analysis based on the Zak phase for the Floquet bands.

The simple model studied here may find a realization in ultracold matter, where the current state of the art allows for manipulations beyond the reach of condensed matter. Previously, the realization of the undriven SSH model in a cold-atom setup permitting the direct measurement of the Zak phase has been demonstrated \cite{Atala2013}. There, the additional driving term could be introduced by modulating the lasers used to produce the dimmerized potential, and the topology of the bands could be characterized by measuring the Zak phase. Furthermore, with the help of the newly developed high-resolution detection and manipulation techniques \cite{Sherson2010,Bakr2010}, one could detect the presence of edge states in finite-size systems. Among all the transitions shown in this manuscript, the detection of those occurring at higher frequencies, such as $\hbar\Omega=4\gamma_0$ (opening of the gap at $\pm\hbar\Omega/2$ containing topological midgap Floquet edge states) and eventually also of the transition at $\hbar\Omega=2\gamma_0$ (annihilation of the native topological states at zero energy), look particularly promising.

\section{Acknowlegdments} 
We acknowledge financial support from SeCyT-UNC. We thank P. M. Perez-Piskunow, H. L. Calvo and G. Usaj for insightful comments on our manuscript. VDL thanks CONICET for the fellowship. LEFFT acknowledges the support of the Abdus Salam International Centre for Theoretical Physics (ICTP, Trieste).

\noindent


\begin{thebibliography}{64}%
\makeatletter
\providecommand \@ifxundefined [1]{%
 \@ifx{#1\undefined}
}%
\providecommand \@ifnum [1]{%
 \ifnum #1\expandafter \@firstoftwo
 \else \expandafter \@secondoftwo
 \fi
}%
\providecommand \@ifx [1]{%
 \ifx #1\expandafter \@firstoftwo
 \else \expandafter \@secondoftwo
 \fi
}%
\providecommand \natexlab [1]{#1}%
\providecommand \enquote  [1]{``#1''}%
\providecommand \bibnamefont  [1]{#1}%
\providecommand \bibfnamefont [1]{#1}%
\providecommand \citenamefont [1]{#1}%
\providecommand \href@noop [0]{\@secondoftwo}%
\providecommand \href [0]{\begingroup \@sanitize@url \@href}%
\providecommand \@href[1]{\@@startlink{#1}\@@href}%
\providecommand \@@href[1]{\endgroup#1\@@endlink}%
\providecommand \@sanitize@url [0]{\catcode `\\12\catcode `\$12\catcode
  `\&12\catcode `\#12\catcode `\^12\catcode `\_12\catcode `\%12\relax}%
\providecommand \@@startlink[1]{}%
\providecommand \@@endlink[0]{}%
\providecommand \url  [0]{\begingroup\@sanitize@url \@url }%
\providecommand \@url [1]{\endgroup\@href {#1}{\urlprefix }}%
\providecommand \urlprefix  [0]{URL }%
\providecommand \Eprint [0]{\href }%
\providecommand \doibase [0]{http://dx.doi.org/}%
\providecommand \selectlanguage [0]{\@gobble}%
\providecommand \bibinfo  [0]{\@secondoftwo}%
\providecommand \bibfield  [0]{\@secondoftwo}%
\providecommand \translation [1]{[#1]}%
\providecommand \BibitemOpen [0]{}%
\providecommand \bibitemStop [0]{}%
\providecommand \bibitemNoStop [0]{.\EOS\space}%
\providecommand \EOS [0]{\spacefactor3000\relax}%
\providecommand \BibitemShut  [1]{\csname bibitem#1\endcsname}%
\let\auto@bib@innerbib\@empty
\bibitem [{\citenamefont {von Klitzing}\ \emph {et~al.}(1980)\citenamefont {von
  Klitzing}, \citenamefont {Dorda},\ and\ \citenamefont
  {Pepper}}]{vonKlitzing1980}%
  \BibitemOpen
  \bibfield  {author} {\bibinfo {author} {\bibfnamefont {K.}~\bibnamefont {von
  Klitzing}}, \bibinfo {author} {\bibfnamefont {G.}~\bibnamefont {Dorda}}, \
  and\ \bibinfo {author} {\bibfnamefont {M.}~\bibnamefont {Pepper}},\
  }\bibfield  {title} {\enquote {\bibinfo {title} {New method for high-accuracy
  determination of the fine-structure constant based on quantized hall
  resistance},}\ }\href {http://link.aps.org/doi/10.1103/PhysRevLett.45.494}
  {\bibfield  {journal} {\bibinfo  {journal} {Phys. Rev. Lett.}\ }\textbf
  {\bibinfo {volume} {45}},\ \bibinfo {pages} {494} (\bibinfo {year}
  {1980})}\BibitemShut {NoStop}%
\bibitem [{\citenamefont {K\"onig}\ \emph {et~al.}(2007)\citenamefont
  {K\"onig}, \citenamefont {Wiedmann}, \citenamefont {Br\"une}, \citenamefont
  {Roth}, \citenamefont {Buhmann}, \citenamefont {Molenkamp}, \citenamefont
  {Qi},\ and\ \citenamefont {Zhang}}]{Koenig2007}%
  \BibitemOpen
  \bibfield  {author} {\bibinfo {author} {\bibfnamefont {M.}~\bibnamefont
  {K\"onig}}, \bibinfo {author} {\bibfnamefont {S.}~\bibnamefont {Wiedmann}},
  \bibinfo {author} {\bibfnamefont {C.}~\bibnamefont {Br\"une}}, \bibinfo
  {author} {\bibfnamefont {A.}~\bibnamefont {Roth}}, \bibinfo {author}
  {\bibfnamefont {H.}~\bibnamefont {Buhmann}}, \bibinfo {author} {\bibfnamefont
  {L.~W.}\ \bibnamefont {Molenkamp}}, \bibinfo {author} {\bibfnamefont {X.-L.}\
  \bibnamefont {Qi}}, \ and\ \bibinfo {author} {\bibfnamefont {S.-C.}\
  \bibnamefont {Zhang}},\ }\bibfield  {title} {\enquote {\bibinfo {title}
  {Quantum spin hall insulator state in hgte quantum wells},}\ }\href
  {http://www.sciencemag.org/content/318/5851/766.abstrac} {\bibfield
  {journal} {\bibinfo  {journal} {Science}\ }\textbf {\bibinfo {volume}
  {318}},\ \bibinfo {pages} {766} (\bibinfo {year} {2007})}\BibitemShut
  {NoStop}%
\bibitem [{\citenamefont {Moore}(2010)}]{Moore2010}%
  \BibitemOpen
  \bibfield  {author} {\bibinfo {author} {\bibfnamefont {J.~E.}\ \bibnamefont
  {Moore}},\ }\bibfield  {title} {\enquote {\bibinfo {title} {The birth of
  topological insulators},}\ }\href {http://dx.doi.org/10.1038/nature08916}
  {\bibfield  {journal} {\bibinfo  {journal} {Nature}\ }\textbf {\bibinfo
  {volume} {464}},\ \bibinfo {pages} {194} (\bibinfo {year}
  {2010})}\BibitemShut {NoStop}%
\bibitem [{\citenamefont {Hasan}\ and\ \citenamefont {Kane}(2010)}]{Hasan2010}%
  \BibitemOpen
  \bibfield  {author} {\bibinfo {author} {\bibfnamefont {M.~Z.}\ \bibnamefont
  {Hasan}}\ and\ \bibinfo {author} {\bibfnamefont {C.~L.}\ \bibnamefont
  {Kane}},\ }\bibfield  {title} {\enquote {\bibinfo {title}
  {\textit{Colloquium} : Topological insulators},}\ }\href {\doibase
  10.1103/RevModPhys.82.3045} {\bibfield  {journal} {\bibinfo  {journal} {Rev.
  Mod. Phys.}\ }\textbf {\bibinfo {volume} {82}},\ \bibinfo {pages} {3045}
  (\bibinfo {year} {2010})}\BibitemShut {NoStop}%
\bibitem [{\citenamefont {Oka}\ and\ \citenamefont {Aoki}(2009)}]{Oka2009}%
  \BibitemOpen
  \bibfield  {author} {\bibinfo {author} {\bibfnamefont {T.}~\bibnamefont
  {Oka}}\ and\ \bibinfo {author} {\bibfnamefont {H.}~\bibnamefont {Aoki}},\
  }\bibfield  {title} {\enquote {\bibinfo {title} {Photovoltaic hall effect in
  graphene},}\ }\href {http://link.aps.org/doi/10.1103/PhysRevB.79.081406}
  {\bibfield  {journal} {\bibinfo  {journal} {Phys. Rev. B}\ }\textbf {\bibinfo
  {volume} {79}},\ \bibinfo {pages} {081406} (\bibinfo {year}
  {2009})}\BibitemShut {NoStop}%
\bibitem [{\citenamefont {Kitagawa}\ \emph {et~al.}(2011)\citenamefont
  {Kitagawa}, \citenamefont {Oka}, \citenamefont {Brataas}, \citenamefont
  {Fu},\ and\ \citenamefont {Demler}}]{Kitagawa2011}%
  \BibitemOpen
  \bibfield  {author} {\bibinfo {author} {\bibfnamefont {T.}~\bibnamefont
  {Kitagawa}}, \bibinfo {author} {\bibfnamefont {T.}~\bibnamefont {Oka}},
  \bibinfo {author} {\bibfnamefont {A.}~\bibnamefont {Brataas}}, \bibinfo
  {author} {\bibfnamefont {L.}~\bibnamefont {Fu}}, \ and\ \bibinfo {author}
  {\bibfnamefont {E.}~\bibnamefont {Demler}},\ }\bibfield  {title} {\enquote
  {\bibinfo {title} {Transport properties of nonequilibrium systems under the
  application of light: Photoinduced quantum hall insulators without landau
  levels},}\ }\href {http://link.aps.org/doi/10.1103/PhysRevB.84.235108}
  {\bibfield  {journal} {\bibinfo  {journal} {Phys. Rev. B}\ }\textbf {\bibinfo
  {volume} {84}},\ \bibinfo {pages} {235108} (\bibinfo {year}
  {2011})}\BibitemShut {NoStop}%
\bibitem [{\citenamefont {Lindner}\ \emph {et~al.}(2011)\citenamefont
  {Lindner}, \citenamefont {Refael},\ and\ \citenamefont
  {Galitski}}]{Lindner2011}%
  \BibitemOpen
  \bibfield  {author} {\bibinfo {author} {\bibfnamefont {N.~H.}\ \bibnamefont
  {Lindner}}, \bibinfo {author} {\bibfnamefont {G.}~\bibnamefont {Refael}}, \
  and\ \bibinfo {author} {\bibfnamefont {V.}~\bibnamefont {Galitski}},\
  }\bibfield  {title} {\enquote {\bibinfo {title} {Floquet topological
  insulator in semiconductor quantum wells},}\ }\href
  {http://dx.doi.org/10.1038/nphys1926} {\bibfield  {journal} {\bibinfo
  {journal} {Nat. Phys.}\ }\textbf {\bibinfo {volume} {7}},\ \bibinfo {pages}
  {490} (\bibinfo {year} {2011})}\BibitemShut {NoStop}%
\bibitem [{\citenamefont {Perez-Piskunow}\ \emph {et~al.}(2014)\citenamefont
  {Perez-Piskunow}, \citenamefont {Usaj}, \citenamefont {Balseiro},\ and\
  \citenamefont {Foa~Torres}}]{Perez-Piskunow2014}%
  \BibitemOpen
  \bibfield  {author} {\bibinfo {author} {\bibfnamefont {P.~M.}\ \bibnamefont
  {Perez-Piskunow}}, \bibinfo {author} {\bibfnamefont {G.}~\bibnamefont
  {Usaj}}, \bibinfo {author} {\bibfnamefont {C.~A.}\ \bibnamefont {Balseiro}},
  \ and\ \bibinfo {author} {\bibfnamefont {L.~E.~F.}\ \bibnamefont
  {Foa~Torres}},\ }\bibfield  {title} {\enquote {\bibinfo {title} {Floquet
  chiral edge states in graphene},}\ }\href {\doibase
  10.1103/PhysRevB.89.121401} {\bibfield  {journal} {\bibinfo  {journal} {Phys.
  Rev. B}\ }\textbf {\bibinfo {volume} {89}},\ \bibinfo {pages} {121401(R)}
  (\bibinfo {year} {2014})}\BibitemShut {NoStop}%
\bibitem [{\citenamefont {Cayssol}\ \emph {et~al.}(2013)\citenamefont
  {Cayssol}, \citenamefont {D\'{o}ra}, \citenamefont {Simon},\ and\
  \citenamefont {Moessner}}]{Cayssol2013}%
  \BibitemOpen
  \bibfield  {author} {\bibinfo {author} {\bibfnamefont {J.}~\bibnamefont
  {Cayssol}}, \bibinfo {author} {\bibfnamefont {B.}~\bibnamefont {D\'{o}ra}},
  \bibinfo {author} {\bibfnamefont {F.}~\bibnamefont {Simon}}, \ and\ \bibinfo
  {author} {\bibfnamefont {R.}~\bibnamefont {Moessner}},\ }\bibfield  {title}
  {\enquote {\bibinfo {title} {Floquet topological insulators},}\ }\href
  {http://dx.doi.org/10.1002/pssr.201206451} {\bibfield  {journal} {\bibinfo
  {journal} {Phys. Status Solidi RRL}\ }\textbf {\bibinfo {volume} {7}},\
  \bibinfo {pages} {101} (\bibinfo {year} {2013})}\BibitemShut {NoStop}%
\bibitem [{\citenamefont {Bukov}\ \emph {et~al.}()\citenamefont {Bukov},
  \citenamefont {D'Alessio},\ and\ \citenamefont {Polkovnikov}}]{Bukov2014}%
  \BibitemOpen
  \bibfield  {author} {\bibinfo {author} {\bibfnamefont {M.}~\bibnamefont
  {Bukov}}, \bibinfo {author} {\bibfnamefont {L.}~\bibnamefont {D'Alessio}}, \
  and\ \bibinfo {author} {\bibfnamefont {A.}~\bibnamefont {Polkovnikov}},\
  }\bibfield  {title} {\enquote {\bibinfo {title} {Universal high-frequency
  behavior of periodically driven systems: from dynamical stabilization to
  floquet engineering},}\ }\href@noop {} {\bibinfo  {journal} {arXiv:1407.4803
  [cond-mat.quant-gas]}\ }\BibitemShut {NoStop}%
\bibitem [{\citenamefont {Tenenbaum~Katan}\ and\ \citenamefont
  {Podolsky}(2013{\natexlab{a}})}]{TenenbaumKatan2013a}%
  \BibitemOpen
\bibfield  {journal} {  }\bibfield  {author} {\bibinfo {author} {\bibfnamefont
  {Y.}~\bibnamefont {Tenenbaum~Katan}}\ and\ \bibinfo {author} {\bibfnamefont
  {D.}~\bibnamefont {Podolsky}},\ }\bibfield  {title} {\enquote {\bibinfo
  {title} {Modulated floquet topological insulators},}\ }\href
  {http://link.aps.org/doi/10.1103/PhysRevLett.110.016802} {\bibfield
  {journal} {\bibinfo  {journal} {Phys. Rev. Lett.}\ }\textbf {\bibinfo
  {volume} {110}},\ \bibinfo {pages} {016802} (\bibinfo {year}
  {2013}{\natexlab{a}})}\BibitemShut {NoStop}%
\bibitem [{\citenamefont {Tenenbaum~Katan}\ and\ \citenamefont
  {Podolsky}(2013{\natexlab{b}})}]{TenenbaumKatan2013}%
  \BibitemOpen
  \bibfield  {author} {\bibinfo {author} {\bibfnamefont {Y.}~\bibnamefont
  {Tenenbaum~Katan}}\ and\ \bibinfo {author} {\bibfnamefont {D.}~\bibnamefont
  {Podolsky}},\ }\bibfield  {title} {\enquote {\bibinfo {title} {Generation and
  manipulation of localized modes in floquet topological insulators},}\ }\href
  {http://link.aps.org/doi/10.1103/PhysRevB.88.224106} {\bibfield  {journal}
  {\bibinfo  {journal} {Phys. Rev. B}\ }\textbf {\bibinfo {volume} {88}},\
  \bibinfo {pages} {224106} (\bibinfo {year} {2013}{\natexlab{b}})}\BibitemShut
  {NoStop}%
\bibitem [{\citenamefont {Su\'arez~Morell}\ and\ \citenamefont
  {Foa~Torres}(2012)}]{SuarezMorell2012}%
  \BibitemOpen
  \bibfield  {author} {\bibinfo {author} {\bibfnamefont {E.}~\bibnamefont
  {Su\'arez~Morell}}\ and\ \bibinfo {author} {\bibfnamefont {L.~E.~F.}\
  \bibnamefont {Foa~Torres}},\ }\bibfield  {title} {\enquote {\bibinfo {title}
  {Radiation effects on the electric properties of bilayer graphene},}\
  }\href@noop {} {\bibfield  {journal} {\bibinfo  {journal} {Phys. Rev. B}\
  }\textbf {\bibinfo {volume} {86}},\ \bibinfo {pages} {125449} (\bibinfo
  {year} {2012})}\BibitemShut {NoStop}%
\bibitem [{\citenamefont {Narayan}(2015)}]{Narayan2015}%
  \BibitemOpen
  \bibfield  {author} {\bibinfo {author} {\bibfnamefont {A.}~\bibnamefont
  {Narayan}},\ }\bibfield  {title} {\enquote {\bibinfo {title} {Floquet
  dynamics in two-dimensional semi-dirac semimetals and three-dimensional dirac
  semimetals},}\ }\href {http://link.aps.org/doi/10.1103/PhysRevB.91.205445}
  {\bibfield  {journal} {\bibinfo  {journal} {Phys. Rev. B}\ }\textbf {\bibinfo
  {volume} {91}},\ \bibinfo {pages} {205445} (\bibinfo {year}
  {2015})}\BibitemShut {NoStop}%
\bibitem [{\citenamefont {Calvo}\ \emph {et~al.}()\citenamefont {Calvo},
  \citenamefont {Foa~Torres}, \citenamefont {Perez-Piskunow}, \citenamefont
  {Balseiro},\ and\ \citenamefont {Usaj}}]{Calvo2015}%
  \BibitemOpen
  \bibfield  {author} {\bibinfo {author} {\bibfnamefont {H.~L.}\ \bibnamefont
  {Calvo}}, \bibinfo {author} {\bibfnamefont {L.~E.~F.}\ \bibnamefont
  {Foa~Torres}}, \bibinfo {author} {\bibfnamefont {P.~M.}\ \bibnamefont
  {Perez-Piskunow}}, \bibinfo {author} {\bibfnamefont {C.~A.}\ \bibnamefont
  {Balseiro}}, \ and\ \bibinfo {author} {\bibfnamefont {G.}~\bibnamefont
  {Usaj}},\ }\bibfield  {title} {\enquote {\bibinfo {title} {Floquet interface
  states in illuminated three dimensional topological insulators},}\
  }\href@noop {} {\bibinfo  {journal} {arXiv:1502.04098 [cond-mat.mes-hall]}\
  }\BibitemShut {NoStop}%
\bibitem [{\citenamefont {Wang}\ \emph {et~al.}(2013)\citenamefont {Wang},
  \citenamefont {Steinberg}, \citenamefont {Jarillo-Herrero},\ and\
  \citenamefont {Gedik}}]{Wang2013}%
  \BibitemOpen
\bibfield  {journal} {  }\bibfield  {author} {\bibinfo {author} {\bibfnamefont
  {Y.~H.}\ \bibnamefont {Wang}}, \bibinfo {author} {\bibfnamefont
  {H.}~\bibnamefont {Steinberg}}, \bibinfo {author} {\bibfnamefont
  {P.}~\bibnamefont {Jarillo-Herrero}}, \ and\ \bibinfo {author} {\bibfnamefont
  {N.}~\bibnamefont {Gedik}},\ }\bibfield  {title} {\enquote {\bibinfo {title}
  {Observation of floquet-bloch states on the surface of a topological
  insulator},}\ }\href {\doibase 10.1126/science.1239834} {\bibfield  {journal}
  {\bibinfo  {journal} {Science}\ }\textbf {\bibinfo {volume} {342}},\ \bibinfo
  {pages} {453} (\bibinfo {year} {2013})}\BibitemShut {NoStop}%
\bibitem [{\citenamefont {Rechtsman}\ \emph {et~al.}(2013)\citenamefont
  {Rechtsman}, \citenamefont {Zeuner}, \citenamefont {Plotnik}, \citenamefont
  {Lumer}, \citenamefont {Podolsky}, \citenamefont {Dreisow}, \citenamefont
  {Nolte}, \citenamefont {Segev},\ and\ \citenamefont
  {Szameit}}]{Rechtsman2013}%
  \BibitemOpen
  \bibfield  {author} {\bibinfo {author} {\bibfnamefont {M.~C.}\ \bibnamefont
  {Rechtsman}}, \bibinfo {author} {\bibfnamefont {J.~M.}\ \bibnamefont
  {Zeuner}}, \bibinfo {author} {\bibfnamefont {Y.}~\bibnamefont {Plotnik}},
  \bibinfo {author} {\bibfnamefont {Y.}~\bibnamefont {Lumer}}, \bibinfo
  {author} {\bibfnamefont {D.}~\bibnamefont {Podolsky}}, \bibinfo {author}
  {\bibfnamefont {F.}~\bibnamefont {Dreisow}}, \bibinfo {author} {\bibfnamefont
  {S.}~\bibnamefont {Nolte}}, \bibinfo {author} {\bibfnamefont
  {M.}~\bibnamefont {Segev}}, \ and\ \bibinfo {author} {\bibfnamefont
  {A.}~\bibnamefont {Szameit}},\ }\bibfield  {title} {\enquote {\bibinfo
  {title} {Photonic floquet topological insulators},}\ }\href
  {http://dx.doi.org/10.1038/nature12066} {\bibfield  {journal} {\bibinfo
  {journal} {Nature}\ }\textbf {\bibinfo {volume} {496}},\ \bibinfo {pages}
  {196} (\bibinfo {year} {2013})}\BibitemShut {NoStop}%
\bibitem [{\citenamefont {Jotzu}\ \emph {et~al.}(2014)\citenamefont {Jotzu},
  \citenamefont {Messer}, \citenamefont {Desbuquois}, \citenamefont {Lebrat},
  \citenamefont {Uehlinger}, \citenamefont {Greif},\ and\ \citenamefont
  {Esslinger}}]{Jotzu2014}%
  \BibitemOpen
  \bibfield  {author} {\bibinfo {author} {\bibfnamefont {G.}~\bibnamefont
  {Jotzu}}, \bibinfo {author} {\bibfnamefont {M.}~\bibnamefont {Messer}},
  \bibinfo {author} {\bibfnamefont {R.}~\bibnamefont {Desbuquois}}, \bibinfo
  {author} {\bibfnamefont {M.}~\bibnamefont {Lebrat}}, \bibinfo {author}
  {\bibfnamefont {T.}~\bibnamefont {Uehlinger}}, \bibinfo {author}
  {\bibfnamefont {D.}~\bibnamefont {Greif}}, \ and\ \bibinfo {author}
  {\bibfnamefont {T.}~\bibnamefont {Esslinger}},\ }\bibfield  {title} {\enquote
  {\bibinfo {title} {Experimental realization of the topological haldane model
  with ultracold fermions},}\ }\href {http://dx.doi.org/10.1038/nature13915}
  {\bibfield  {journal} {\bibinfo  {journal} {Nature}\ }\textbf {\bibinfo
  {volume} {515}},\ \bibinfo {pages} {237} (\bibinfo {year}
  {2014})}\BibitemShut {NoStop}%
\bibitem [{\citenamefont {Gao}\ \emph {et~al.}()\citenamefont {Gao},
  \citenamefont {Gao}, \citenamefont {Shi}, \citenamefont {Yang}, \citenamefont
  {Lin}, \citenamefont {Joannopoulos}, \citenamefont {Soljacic}, \citenamefont
  {Chen}, \citenamefont {Lu}, \citenamefont {Chong},\ and\ \citenamefont
  {Zhang}}]{Gao2015}%
  \BibitemOpen
  \bibfield  {author} {\bibinfo {author} {\bibfnamefont {F.}~\bibnamefont
  {Gao}}, \bibinfo {author} {\bibfnamefont {Z.}~\bibnamefont {Gao}}, \bibinfo
  {author} {\bibfnamefont {X.}~\bibnamefont {Shi}}, \bibinfo {author}
  {\bibfnamefont {Z.}~\bibnamefont {Yang}}, \bibinfo {author} {\bibfnamefont
  {X.}~\bibnamefont {Lin}}, \bibinfo {author} {\bibfnamefont {J.~D.}\
  \bibnamefont {Joannopoulos}}, \bibinfo {author} {\bibfnamefont
  {M.}~\bibnamefont {Soljacic}}, \bibinfo {author} {\bibfnamefont
  {H.}~\bibnamefont {Chen}}, \bibinfo {author} {\bibfnamefont {L.}~\bibnamefont
  {Lu}}, \bibinfo {author} {\bibfnamefont {Y.}~\bibnamefont {Chong}}, \ and\
  \bibinfo {author} {\bibfnamefont {B.}~\bibnamefont {Zhang}},\ }\bibfield
  {title} {\enquote {\bibinfo {title} {Probing the limits of topological
  protection in a designer surface plasmon structure},}\ }\href
  {http://arxiv.org/abs/1504.07809} {\bibinfo  {journal} {arXiv:1504.07809
  [physics.optics]}\ }\BibitemShut {NoStop}%
\bibitem [{\citenamefont {Usaj}\ \emph {et~al.}(2014)\citenamefont {Usaj},
  \citenamefont {Perez-Piskunow}, \citenamefont {Foa~Torres},\ and\
  \citenamefont {Balseiro}}]{Usaj2014}%
  \BibitemOpen
\bibfield  {journal} {  }\bibfield  {author} {\bibinfo {author} {\bibfnamefont
  {G.}~\bibnamefont {Usaj}}, \bibinfo {author} {\bibfnamefont {P.~M.}\
  \bibnamefont {Perez-Piskunow}}, \bibinfo {author} {\bibfnamefont {L.~E.~F.}\
  \bibnamefont {Foa~Torres}}, \ and\ \bibinfo {author} {\bibfnamefont {C.~A.}\
  \bibnamefont {Balseiro}},\ }\bibfield  {title} {\enquote {\bibinfo {title}
  {Irradiated graphene as a tunable floquet topological insulator},}\ }\href
  {\doibase 10.1103/PhysRevB.90.115423} {\bibfield  {journal} {\bibinfo
  {journal} {Phys. Rev. B}\ }\textbf {\bibinfo {volume} {90}},\ \bibinfo
  {pages} {115423} (\bibinfo {year} {2014})}\BibitemShut {NoStop}%
\bibitem [{\citenamefont {Dahlhaus}\ \emph {et~al.}()\citenamefont {Dahlhaus},
  \citenamefont {Fregoso},\ and\ \citenamefont {Moore}}]{Dahlhaus2014}%
  \BibitemOpen
  \bibfield  {author} {\bibinfo {author} {\bibfnamefont {J.~P.}\ \bibnamefont
  {Dahlhaus}}, \bibinfo {author} {\bibfnamefont {B.~M.}\ \bibnamefont
  {Fregoso}}, \ and\ \bibinfo {author} {\bibfnamefont {J.~E.}\ \bibnamefont
  {Moore}},\ }\bibfield  {title} {\enquote {\bibinfo {title} {Magnetization
  signatures of light-induced quantum hall edge states},}\ }\href {\doibase
  http://arxiv.org/abs/1408.6811} {\bibfield  {journal} {\bibinfo  {journal}
  {arXiv:1408.6811 [cond-mat.mes-hall]}\
  }http://arxiv.org/abs/1408.6811}\BibitemShut {NoStop}%
\bibitem [{\citenamefont {Thakurathi}\ \emph {et~al.}(2013)\citenamefont
  {Thakurathi}, \citenamefont {Patel}, \citenamefont {Sen},\ and\ \citenamefont
  {Dutta}}]{Thakurathi2013}%
  \BibitemOpen
  \bibfield  {author} {\bibinfo {author} {\bibfnamefont {M.}~\bibnamefont
  {Thakurathi}}, \bibinfo {author} {\bibfnamefont {A.~A.}\ \bibnamefont
  {Patel}}, \bibinfo {author} {\bibfnamefont {D.}~\bibnamefont {Sen}}, \ and\
  \bibinfo {author} {\bibfnamefont {A.}~\bibnamefont {Dutta}},\ }\bibfield
  {title} {\enquote {\bibinfo {title} {Floquet generation of majorana end modes
  and topological invariants},}\ }\href
  {http://link.aps.org/doi/10.1103/PhysRevB.88.155133} {\bibfield  {journal}
  {\bibinfo  {journal} {Phys. Rev. B}\ }\textbf {\bibinfo {volume} {88}},\
  \bibinfo {pages} {155133} (\bibinfo {year} {2013})}\BibitemShut {NoStop}%
\bibitem [{\citenamefont {Sentef}\ \emph {et~al.}(2015)\citenamefont {Sentef},
  \citenamefont {Claassen}, \citenamefont {Kemper}, \citenamefont {Moritz},
  \citenamefont {Oka}, \citenamefont {Freericks},\ and\ \citenamefont
  {Devereaux}}]{Sentef2015}%
  \BibitemOpen
  \bibfield  {author} {\bibinfo {author} {\bibfnamefont {M.}~\bibnamefont
  {Sentef}}, \bibinfo {author} {\bibfnamefont {M.}~\bibnamefont {Claassen}},
  \bibinfo {author} {\bibfnamefont {A.}~\bibnamefont {Kemper}}, \bibinfo
  {author} {\bibfnamefont {B.}~\bibnamefont {Moritz}}, \bibinfo {author}
  {\bibfnamefont {T.}~\bibnamefont {Oka}}, \bibinfo {author} {\bibfnamefont
  {J.}~\bibnamefont {Freericks}}, \ and\ \bibinfo {author} {\bibfnamefont
  {T.}~\bibnamefont {Devereaux}},\ }\bibfield  {title} {\enquote {\bibinfo
  {title} {Theory of floquet band formation and local pseudospin textures in
  pump-probe photoemission of graphene},}\ }\href
  {http://dx.doi.org/10.1038/ncomms8047} {\bibfield  {journal} {\bibinfo
  {journal} {Nat. Comms.}\ }\textbf {\bibinfo {volume} {6}},\  (\bibinfo {year}
  {2015})}\BibitemShut {NoStop}%
\bibitem [{\citenamefont {Rudner}\ \emph {et~al.}(2013)\citenamefont {Rudner},
  \citenamefont {Lindner}, \citenamefont {Berg},\ and\ \citenamefont
  {Levin}}]{Rudner2013}%
  \BibitemOpen
  \bibfield  {author} {\bibinfo {author} {\bibfnamefont {M.~S.}\ \bibnamefont
  {Rudner}}, \bibinfo {author} {\bibfnamefont {N.~H.}\ \bibnamefont {Lindner}},
  \bibinfo {author} {\bibfnamefont {E.}~\bibnamefont {Berg}}, \ and\ \bibinfo
  {author} {\bibfnamefont {M.}~\bibnamefont {Levin}},\ }\bibfield  {title}
  {\enquote {\bibinfo {title} {Anomalous edge states and the bulk-edge
  correspondence for periodically-driven two dimensional systems},}\
  }\href@noop {} {\bibfield  {journal} {\bibinfo  {journal} {Phys. Rev. X}\
  }\textbf {\bibinfo {volume} {3}},\ \bibinfo {pages} {031005} (\bibinfo {year}
  {2013})}\BibitemShut {NoStop}%
\bibitem [{\citenamefont {Ho}\ and\ \citenamefont {Gong}(2014)}]{Ho2014}%
  \BibitemOpen
  \bibfield  {author} {\bibinfo {author} {\bibfnamefont {D.~Y.}\ \bibnamefont
  {Ho}}\ and\ \bibinfo {author} {\bibfnamefont {J.}~\bibnamefont {Gong}},\
  }\bibfield  {title} {\enquote {\bibinfo {title} {Topological effects in
  chiral symmetric driven systems},}\ }\href {\doibase
  http://dx.doi.org/10.1103/PhysRevB.90.195419} {\bibfield  {journal} {\bibinfo
   {journal} {Phys. Rev. B}\ }\textbf {\bibinfo {volume} {90}},\ \bibinfo
  {pages} {195419} (\bibinfo {year} {2014})}\BibitemShut {NoStop}%
\bibitem [{\citenamefont {Carpentier}\ \emph {et~al.}(2015)\citenamefont
  {Carpentier}, \citenamefont {Delplace}, \citenamefont {Fruchart},\ and\
  \citenamefont {Gawędzki}}]{Carpentier2015}%
  \BibitemOpen
  \bibfield  {author} {\bibinfo {author} {\bibfnamefont {D.}~\bibnamefont
  {Carpentier}}, \bibinfo {author} {\bibfnamefont {P.}~\bibnamefont
  {Delplace}}, \bibinfo {author} {\bibfnamefont {M.}~\bibnamefont {Fruchart}},
  \ and\ \bibinfo {author} {\bibfnamefont {K.}~\bibnamefont {Gawędzki}},\
  }\bibfield  {title} {\enquote {\bibinfo {title} {Topological index for
  periodically driven time-reversal invariant 2d systems},}\ }\href
  {http://link.aps.org/doi/10.1103/PhysRevLett.114.106806} {\bibfield
  {journal} {\bibinfo  {journal} {Phys. Rev. Lett.}\ }\textbf {\bibinfo
  {volume} {114}},\ \bibinfo {pages} {106806} (\bibinfo {year}
  {2015})}\BibitemShut {NoStop}%
\bibitem [{\citenamefont {Goldman}\ and\ \citenamefont
  {Dalibard}(2014)}]{Goldman2014}%
  \BibitemOpen
  \bibfield  {author} {\bibinfo {author} {\bibfnamefont {N.}~\bibnamefont
  {Goldman}}\ and\ \bibinfo {author} {\bibfnamefont {J.}~\bibnamefont
  {Dalibard}},\ }\bibfield  {title} {\enquote {\bibinfo {title} {Periodically
  driven quantum systems: Effective hamiltonians and engineered gauge
  fields},}\ }\href {http://link.aps.org/doi/10.1103/PhysRevX.4.031027}
  {\bibfield  {journal} {\bibinfo  {journal} {Phys. Rev. X}\ }\textbf {\bibinfo
  {volume} {4}},\ \bibinfo {pages} {031027} (\bibinfo {year}
  {2014})}\BibitemShut {NoStop}%
\bibitem [{\citenamefont {D'Alessio}\ and\ \citenamefont
  {Rigol}()}]{DAlessio2014}%
  \BibitemOpen
  \bibfield  {author} {\bibinfo {author} {\bibfnamefont {L.}~\bibnamefont
  {D'Alessio}}\ and\ \bibinfo {author} {\bibfnamefont {M.}~\bibnamefont
  {Rigol}},\ }\bibfield  {title} {\enquote {\bibinfo {title} {Dynamical
  preparation of floquet chern insulators: A no-go theorem and the
  experiments},}\ }\href {http://arxiv.org/abs/1409.6319} {\bibinfo  {journal}
  {arXiv:1409.6319 [cond-mat.quant-gas]}\ }\BibitemShut {NoStop}%
\bibitem [{\citenamefont {Seetharam}\ \emph {et~al.}()\citenamefont
  {Seetharam}, \citenamefont {Bardyn}, \citenamefont {Lindner}, \citenamefont
  {Rudner},\ and\ \citenamefont {Refael}}]{Seetharam2015}%
  \BibitemOpen
\bibfield  {journal} {  }\bibfield  {author} {\bibinfo {author} {\bibfnamefont
  {K.~I.}\ \bibnamefont {Seetharam}}, \bibinfo {author} {\bibfnamefont {C.-E.}\
  \bibnamefont {Bardyn}}, \bibinfo {author} {\bibfnamefont {N.~H.}\
  \bibnamefont {Lindner}}, \bibinfo {author} {\bibfnamefont {M.~S.}\
  \bibnamefont {Rudner}}, \ and\ \bibinfo {author} {\bibfnamefont
  {G.}~\bibnamefont {Refael}},\ }\bibfield  {title} {\enquote {\bibinfo {title}
  {Controlled population of floquet-bloch states via coupling to bose and fermi
  baths},}\ }\href@noop {} {\bibinfo  {journal} {arXiv:1502.02664
  [cond-mat.mes-hall]}\ }\BibitemShut {NoStop}%
\bibitem [{\citenamefont {Iadecola}\ \emph {et~al.}()\citenamefont {Iadecola},
  \citenamefont {Neupert},\ and\ \citenamefont {Chamon}}]{Iadecola2015}%
  \BibitemOpen
\bibfield  {journal} {  }\bibfield  {author} {\bibinfo {author} {\bibfnamefont
  {T.}~\bibnamefont {Iadecola}}, \bibinfo {author} {\bibfnamefont
  {T.}~\bibnamefont {Neupert}}, \ and\ \bibinfo {author} {\bibfnamefont
  {C.}~\bibnamefont {Chamon}},\ }\bibfield  {title} {\enquote {\bibinfo {title}
  {Occupation of topological floquet bands in open systems},}\ }\href@noop {}
  {\bibinfo  {journal} {arXiv:1502.05047 [cond-mat.mes-hall]}\ }\BibitemShut
  {NoStop}%
\bibitem [{\citenamefont {Kundu}\ \emph {et~al.}(2014)\citenamefont {Kundu},
  \citenamefont {Fertig},\ and\ \citenamefont {Seradjeh}}]{Kundu2014}%
  \BibitemOpen
\bibfield  {journal} {  }\bibfield  {author} {\bibinfo {author} {\bibfnamefont
  {A.}~\bibnamefont {Kundu}}, \bibinfo {author} {\bibfnamefont {H.~A.}\
  \bibnamefont {Fertig}}, \ and\ \bibinfo {author} {\bibfnamefont
  {B.}~\bibnamefont {Seradjeh}},\ }\bibfield  {title} {\enquote {\bibinfo
  {title} {Effective theory of floquet topological transitions},}\ }\href@noop
  {} {\bibfield  {journal} {\bibinfo  {journal} {Phys. Rev. Lett.}\ }\textbf
  {\bibinfo {volume} {113}},\ \bibinfo {pages} {236803} (\bibinfo {year}
  {2014})}\BibitemShut {NoStop}%
\bibitem [{\citenamefont {Dehghani}\ \emph {et~al.}()\citenamefont {Dehghani},
  \citenamefont {Oka},\ and\ \citenamefont {Mitra}}]{Dehghani2014}%
  \BibitemOpen
  \bibfield  {author} {\bibinfo {author} {\bibfnamefont {H.}~\bibnamefont
  {Dehghani}}, \bibinfo {author} {\bibfnamefont {T.}~\bibnamefont {Oka}}, \
  and\ \bibinfo {author} {\bibfnamefont {A.}~\bibnamefont {Mitra}},\ }\bibfield
   {title} {\enquote {\bibinfo {title} {Dissipative floquet topological
  systems},}\ }\href@noop {} {\bibinfo  {journal} {arXiv:1406.6626
  [cond-mat.mes-hall]}\ }\BibitemShut {NoStop}%
\bibitem [{\citenamefont {Farrell}\ and\ \citenamefont
  {Pereg-Barnea}()}]{Farrell2015}%
  \BibitemOpen
\bibfield  {journal} {  }\bibfield  {author} {\bibinfo {author} {\bibfnamefont
  {A.}~\bibnamefont {Farrell}}\ and\ \bibinfo {author} {\bibfnamefont
  {T.}~\bibnamefont {Pereg-Barnea}},\ }\bibfield  {title} {\enquote {\bibinfo
  {title} {Edge state transport in floquet topological insulators},}\
  }\href@noop {} {\bibinfo  {journal} {arXiv:1505.05584 [cond-mat.str-el]}\
  }\BibitemShut {NoStop}%
\bibitem [{\citenamefont {Dehghani}\ \emph {et~al.}(2015)\citenamefont
  {Dehghani}, \citenamefont {Oka},\ and\ \citenamefont {Mitra}}]{Dehghani2015}%
  \BibitemOpen
\bibfield  {journal} {  }\bibfield  {author} {\bibinfo {author} {\bibfnamefont
  {H.}~\bibnamefont {Dehghani}}, \bibinfo {author} {\bibfnamefont
  {T.}~\bibnamefont {Oka}}, \ and\ \bibinfo {author} {\bibfnamefont
  {A.}~\bibnamefont {Mitra}},\ }\bibfield  {title} {\enquote {\bibinfo {title}
  {Out-of-equilibrium electrons and the hall conductance of a floquet
  topological insulator},}\ }\href
  {http://link.aps.org/doi/10.1103/PhysRevB.91.155422} {\bibfield  {journal}
  {\bibinfo  {journal} {Phys. Rev. B}\ }\textbf {\bibinfo {volume} {91}},\
  \bibinfo {pages} {155422} (\bibinfo {year} {2015})}\BibitemShut {NoStop}%
\bibitem [{\citenamefont {Foa~Torres}\ \emph {et~al.}(2014)\citenamefont
  {Foa~Torres}, \citenamefont {Perez-Piskunow}, \citenamefont {Balseiro},\ and\
  \citenamefont {Usaj}}]{Foa2014}%
  \BibitemOpen
  \bibfield  {author} {\bibinfo {author} {\bibfnamefont {L.~E.~F.}\
  \bibnamefont {Foa~Torres}}, \bibinfo {author} {\bibfnamefont {P.~M.}\
  \bibnamefont {Perez-Piskunow}}, \bibinfo {author} {\bibfnamefont {C.~A.}\
  \bibnamefont {Balseiro}}, \ and\ \bibinfo {author} {\bibfnamefont
  {G.}~\bibnamefont {Usaj}},\ }\bibfield  {title} {\enquote {\bibinfo {title}
  {Multiterminal conductance of a floquet topological insulator},}\ }\href
  {http://journals.aps.org/prl/abstract/10.1103/PhysRevLett.113.266801}
  {\bibfield  {journal} {\bibinfo  {journal} {Phys. Rev. Lett.}\ }\textbf
  {\bibinfo {volume} {113}},\ \bibinfo {pages} {266801} (\bibinfo {year}
  {2014})}\BibitemShut {NoStop}%
\bibitem [{\citenamefont {Thouless}\ \emph {et~al.}(1982)\citenamefont
  {Thouless}, \citenamefont {Kohmoto}, \citenamefont {Nightingale},\ and\
  \citenamefont {den Nijs}}]{Thouless1982}%
  \BibitemOpen
  \bibfield  {author} {\bibinfo {author} {\bibfnamefont {D.~J.}\ \bibnamefont
  {Thouless}}, \bibinfo {author} {\bibfnamefont {M.}~\bibnamefont {Kohmoto}},
  \bibinfo {author} {\bibfnamefont {M.~P.}\ \bibnamefont {Nightingale}}, \ and\
  \bibinfo {author} {\bibfnamefont {M.}~\bibnamefont {den Nijs}},\ }\bibfield
  {title} {\enquote {\bibinfo {title} {Quantized hall conductance in a
  two-dimensional periodic potential},}\ }\href
  {http://link.aps.org/doi/10.1103/PhysRevLett.49.405} {\bibfield  {journal}
  {\bibinfo  {journal} {Phys. Rev. Lett.}\ }\textbf {\bibinfo {volume} {49}},\
  \bibinfo {pages} {405} (\bibinfo {year} {1982})}\BibitemShut {NoStop}%
\bibitem [{\citenamefont {Su}\ \emph {et~al.}(1979)\citenamefont {Su},
  \citenamefont {Schrieffer},\ and\ \citenamefont {Heeger}}]{Su1979}%
  \BibitemOpen
  \bibfield  {author} {\bibinfo {author} {\bibfnamefont {W.~P.}\ \bibnamefont
  {Su}}, \bibinfo {author} {\bibfnamefont {J.~R.}\ \bibnamefont {Schrieffer}},
  \ and\ \bibinfo {author} {\bibfnamefont {A.~J.}\ \bibnamefont {Heeger}},\
  }\bibfield  {title} {\enquote {\bibinfo {title} {Solitons in
  polyacetylene},}\ }\href {http://link.aps.org/abstract/PRL/v42/p1698}
  {\bibfield  {journal} {\bibinfo  {journal} {Phys. Rev. Lett.}\ }\textbf
  {\bibinfo {volume} {42}},\ \bibinfo {pages} {1698} (\bibinfo {year}
  {1979})}\BibitemShut {NoStop}%
\bibitem [{\citenamefont {Heeger}\ \emph {et~al.}(1988)\citenamefont {Heeger},
  \citenamefont {Kivelson}, \citenamefont {Schrieffer},\ and\ \citenamefont
  {Su}}]{Heeger1988}%
  \BibitemOpen
  \bibfield  {author} {\bibinfo {author} {\bibfnamefont {A.}~\bibnamefont
  {Heeger}}, \bibinfo {author} {\bibfnamefont {S.}~\bibnamefont {Kivelson}},
  \bibinfo {author} {\bibfnamefont {J.}~\bibnamefont {Schrieffer}}, \ and\
  \bibinfo {author} {\bibfnamefont {W.}~\bibnamefont {Su}},\ }\bibfield
  {title} {\enquote {\bibinfo {title} {Solitons in conducting polymers},}\
  }\href {http://link.aps.org/doi/10.1103/RevModPhys.60.781} {\bibfield
  {journal} {\bibinfo  {journal} {Rev. Mod. Phys.}\ }\textbf {\bibinfo {volume}
  {60}},\ \bibinfo {pages} {781} (\bibinfo {year} {1988})}\BibitemShut
  {NoStop}%
\bibitem [{\citenamefont {Fefferman}\ \emph {et~al.}(2014)\citenamefont
  {Fefferman}, \citenamefont {Lee-Thorp},\ and\ \citenamefont
  {Weinstein}}]{Fefferman2014}%
  \BibitemOpen
  \bibfield  {author} {\bibinfo {author} {\bibfnamefont {C.~L.}\ \bibnamefont
  {Fefferman}}, \bibinfo {author} {\bibfnamefont {J.~P.}\ \bibnamefont
  {Lee-Thorp}}, \ and\ \bibinfo {author} {\bibfnamefont {M.~I.}\ \bibnamefont
  {Weinstein}},\ }\bibfield  {title} {\enquote {\bibinfo {title} {Topologically
  protected states in one-dimensional continuous systems and dirac points},}\
  }\href {\doibase 10.1073/pnas.1407391111} {\bibfield  {journal} {\bibinfo
  {journal} {PNAS}\ }\textbf {\bibinfo {volume} {111}},\ \bibinfo {pages}
  {8759} (\bibinfo {year} {2014})}\BibitemShut {NoStop}%
\bibitem [{\citenamefont {Li}\ \emph {et~al.}(2014)\citenamefont {Li},
  \citenamefont {Xu},\ and\ \citenamefont {Chen}}]{Li2014}%
  \BibitemOpen
  \bibfield  {author} {\bibinfo {author} {\bibfnamefont {L.}~\bibnamefont
  {Li}}, \bibinfo {author} {\bibfnamefont {Z.}~\bibnamefont {Xu}}, \ and\
  \bibinfo {author} {\bibfnamefont {S.}~\bibnamefont {Chen}},\ }\bibfield
  {title} {\enquote {\bibinfo {title} {Topological phases of generalized
  su-schrieffer-heeger model},}\ }\href
  {http://link.aps.org/doi/10.1103/PhysRevB.89.085111} {\bibfield  {journal}
  {\bibinfo  {journal} {Phys. Rev. B}\ }\textbf {\bibinfo {volume} {89}},\
  \bibinfo {pages} {085111} (\bibinfo {year} {2014})}\BibitemShut {NoStop}%
\bibitem [{\citenamefont {Asboth}\ \emph {et~al.}(2014)\citenamefont {Asboth},
  \citenamefont {Tarasinski},\ and\ \citenamefont {Delplace}}]{Asboth2014}%
  \BibitemOpen
  \bibfield  {author} {\bibinfo {author} {\bibfnamefont {J.~K.}\ \bibnamefont
  {Asboth}}, \bibinfo {author} {\bibfnamefont {B.}~\bibnamefont {Tarasinski}},
  \ and\ \bibinfo {author} {\bibfnamefont {P.}~\bibnamefont {Delplace}},\
  }\bibfield  {title} {\enquote {\bibinfo {title} {Chiral symmetry and
  bulk--boundary correspondence in periodically driven one-dimensional
  systems},}\ }\href {\doibase http://dx.doi.org/10.1103/PhysRevB.90.125143}
  {\bibfield  {journal} {\bibinfo  {journal} {Phys. Rev. B}\ }\textbf {\bibinfo
  {volume} {90}},\ \bibinfo {pages} {125143} (\bibinfo {year}
  {2014})}\BibitemShut {NoStop}%
\bibitem [{\citenamefont {Zak}(1989)}]{Zak1989}%
  \BibitemOpen
  \bibfield  {author} {\bibinfo {author} {\bibfnamefont {J.}~\bibnamefont
  {Zak}},\ }\bibfield  {title} {\enquote {\bibinfo {title} {Berry’s phase for
  energy bands in solids},}\ }\href
  {http://link.aps.org/doi/10.1103/PhysRevLett.62.2747} {\bibfield  {journal}
  {\bibinfo  {journal} {Phys. Rev. Lett.}\ }\textbf {\bibinfo {volume} {62}},\
  \bibinfo {pages} {2747} (\bibinfo {year} {1989})}\BibitemShut {NoStop}%
\bibitem [{\citenamefont {Choudhury}\ and\ \citenamefont
  {Mueller}(2014)}]{Choudhury2014}%
  \BibitemOpen
  \bibfield  {author} {\bibinfo {author} {\bibfnamefont {S.}~\bibnamefont
  {Choudhury}}\ and\ \bibinfo {author} {\bibfnamefont {E.~J.}\ \bibnamefont
  {Mueller}},\ }\bibfield  {title} {\enquote {\bibinfo {title} {Stability of a
  floquet bose-einstein condensate in a one-dimensional optical lattice},}\
  }\href {\doibase http://dx.doi.org/10.1103/PhysRevA.90.013621} {\bibfield
  {journal} {\bibinfo  {journal} {Phys. Rev. A}\ }\textbf {\bibinfo {volume}
  {90}},\ \bibinfo {pages} {013621} (\bibinfo {year} {2014})}\BibitemShut
  {NoStop}%
\bibitem [{\citenamefont {Goldman}\ \emph {et~al.}(2015)\citenamefont
  {Goldman}, \citenamefont {Dalibard}, \citenamefont {Aidelsburger},\ and\
  \citenamefont {Cooper}}]{Goldman2015}%
  \BibitemOpen
  \bibfield  {author} {\bibinfo {author} {\bibfnamefont {N.}~\bibnamefont
  {Goldman}}, \bibinfo {author} {\bibfnamefont {J.}~\bibnamefont {Dalibard}},
  \bibinfo {author} {\bibfnamefont {M.}~\bibnamefont {Aidelsburger}}, \ and\
  \bibinfo {author} {\bibfnamefont {N.~R.}\ \bibnamefont {Cooper}},\ }\bibfield
   {title} {\enquote {\bibinfo {title} {Periodically driven quantum matter: The
  case of resonant modulations},}\ }\href
  {http://link.aps.org/doi/10.1103/PhysRevA.91.033632} {\bibfield  {journal}
  {\bibinfo  {journal} {Phys. Rev. A}\ }\textbf {\bibinfo {volume} {91}},\
  \bibinfo {pages} {033632} (\bibinfo {year} {2015})}\BibitemShut {NoStop}%
\bibitem [{\citenamefont {Quelle}\ \emph {et~al.}(2015)\citenamefont {Quelle},
  \citenamefont {Goerbig},\ and\ \citenamefont {Morais~Smith}}]{Quelle2015}%
  \BibitemOpen
  \bibfield  {author} {\bibinfo {author} {\bibfnamefont {A.}~\bibnamefont
  {Quelle}}, \bibinfo {author} {\bibfnamefont {M.}~\bibnamefont {Goerbig}}, \
  and\ \bibinfo {author} {\bibfnamefont {C.}~\bibnamefont {Morais~Smith}},\
  }\bibfield  {title} {\enquote {\bibinfo {title} {Artificial graphene under
  the spotlight: a realisation of the bernevig-hughes-zhang model},}\
  }\href@noop {} {\bibfield  {journal} {\bibinfo  {journal} {arXiv:1503.02635
  [cond-mat.str-el]}\ } (\bibinfo {year} {2015})}\BibitemShut {NoStop}%
\bibitem [{\citenamefont {Atala}\ \emph {et~al.}(2013)\citenamefont {Atala},
  \citenamefont {Aidelsburger}, \citenamefont {Barreiro}, \citenamefont
  {Abanin}, \citenamefont {Kitagawa}, \citenamefont {Demler},\ and\
  \citenamefont {Bloch}}]{Atala2013}%
  \BibitemOpen
  \bibfield  {author} {\bibinfo {author} {\bibfnamefont {M.}~\bibnamefont
  {Atala}}, \bibinfo {author} {\bibfnamefont {M.}~\bibnamefont {Aidelsburger}},
  \bibinfo {author} {\bibfnamefont {J.~T.}\ \bibnamefont {Barreiro}}, \bibinfo
  {author} {\bibfnamefont {D.}~\bibnamefont {Abanin}}, \bibinfo {author}
  {\bibfnamefont {T.}~\bibnamefont {Kitagawa}}, \bibinfo {author}
  {\bibfnamefont {E.}~\bibnamefont {Demler}}, \ and\ \bibinfo {author}
  {\bibfnamefont {I.}~\bibnamefont {Bloch}},\ }\bibfield  {title} {\enquote
  {\bibinfo {title} {Direct measurement of the zak phase in topological bloch
  bands},}\ }\href {http://dx.doi.org/10.1038/nphys2790} {\bibfield  {journal}
  {\bibinfo  {journal} {Nat. Phys.}\ }\textbf {\bibinfo {volume} {9}},\
  \bibinfo {pages} {795} (\bibinfo {year} {2013})}\BibitemShut {NoStop}%
\bibitem [{\citenamefont {Aidelsburger}\ \emph {et~al.}(2015)\citenamefont
  {Aidelsburger}, \citenamefont {Lohse}, \citenamefont {Schweizer},
  \citenamefont {Atala}, \citenamefont {Barreiro}, \citenamefont {Nascimbene},
  \citenamefont {Cooper}, \citenamefont {Bloch},\ and\ \citenamefont
  {Goldman}}]{Aidelsburger2015}%
  \BibitemOpen
  \bibfield  {author} {\bibinfo {author} {\bibfnamefont {M.}~\bibnamefont
  {Aidelsburger}}, \bibinfo {author} {\bibfnamefont {M.}~\bibnamefont {Lohse}},
  \bibinfo {author} {\bibfnamefont {C.}~\bibnamefont {Schweizer}}, \bibinfo
  {author} {\bibfnamefont {M.}~\bibnamefont {Atala}}, \bibinfo {author}
  {\bibfnamefont {J.~T.}\ \bibnamefont {Barreiro}}, \bibinfo {author}
  {\bibfnamefont {S.}~\bibnamefont {Nascimbene}}, \bibinfo {author}
  {\bibfnamefont {N.~R.}\ \bibnamefont {Cooper}}, \bibinfo {author}
  {\bibfnamefont {I.}~\bibnamefont {Bloch}}, \ and\ \bibinfo {author}
  {\bibfnamefont {N.}~\bibnamefont {Goldman}},\ }\bibfield  {title} {\enquote
  {\bibinfo {title} {Measuring the chern number of hofstadter bands with
  ultracold bosonic atoms},}\ }\href {http://dx.doi.org/10.1038/nphys3171}
  {\bibfield  {journal} {\bibinfo  {journal} {Nat. Phys.}\ }\textbf {\bibinfo
  {volume} {11}},\ \bibinfo {pages} {162} (\bibinfo {year} {2015})}\BibitemShut
  {NoStop}%
\bibitem [{\citenamefont {Peierls}(1955)}]{Peierls1955}%
  \BibitemOpen
  \bibfield  {author} {\bibinfo {author} {\bibfnamefont {R.}~\bibnamefont
  {Peierls}},\ }\href@noop {} {\emph {\bibinfo {title} {Quantum Theory of
  Solids}}}\ (\bibinfo  {publisher} {Clarendon, Oxford},\ \bibinfo {year}
  {1955})\BibitemShut {NoStop}%
\bibitem [{\citenamefont {Peierls}(1991)}]{PeierlsMoreSurprises}%
  \BibitemOpen
  \bibfield  {author} {\bibinfo {author} {\bibfnamefont {R.}~\bibnamefont
  {Peierls}},\ }\href@noop {} {\emph {\bibinfo {title} {More Surprises in
  Theoretical Physics}}}\ (\bibinfo  {publisher} {Princeton University Press},\
  \bibinfo {year} {1991})\BibitemShut {NoStop}%
\bibitem [{\citenamefont {Franz}\ and\ \citenamefont
  {Molenkamp}(2013)}]{Franz2013}%
  \BibitemOpen
  \bibinfo {editor} {\bibfnamefont {M.}~\bibnamefont {Franz}}\ and\ \bibinfo
  {editor} {\bibfnamefont {L.}~\bibnamefont {Molenkamp}},\ eds.,\ \href@noop {}
  {\emph {\bibinfo {title} {Topological Insulators}}}\ (\bibinfo  {publisher}
  {Elsevier},\ \bibinfo {year} {2013})\BibitemShut {NoStop}%
\bibitem [{\citenamefont {Grifoni}\ and\ \citenamefont
  {H\"anggi}(1998)}]{Grifoni1998}%
  \BibitemOpen
  \bibfield  {author} {\bibinfo {author} {\bibfnamefont {M.}~\bibnamefont
  {Grifoni}}\ and\ \bibinfo {author} {\bibfnamefont {P.}~\bibnamefont
  {H\"anggi}},\ }\bibfield  {title} {\enquote {\bibinfo {title} {Driven quantum
  tunneling},}\ }\href {\doibase
  http://dx.doi.org/10.1016/S0370-1573(98)00022-2} {\bibfield  {journal}
  {\bibinfo  {journal} {Physics Reports}\ }\textbf {\bibinfo {volume} {304}},\
  \bibinfo {pages} {229 } (\bibinfo {year} {1998})}\BibitemShut {NoStop}%
\bibitem [{\citenamefont {Kohler}\ \emph {et~al.}(2005)\citenamefont {Kohler},
  \citenamefont {Lehmann},\ and\ \citenamefont {H\"anggi}}]{Kohler2005}%
  \BibitemOpen
  \bibfield  {author} {\bibinfo {author} {\bibfnamefont {S.}~\bibnamefont
  {Kohler}}, \bibinfo {author} {\bibfnamefont {J.}~\bibnamefont {Lehmann}}, \
  and\ \bibinfo {author} {\bibfnamefont {P.}~\bibnamefont {H\"anggi}},\
  }\bibfield  {title} {\enquote {\bibinfo {title} {Driven quantum transport on
  the nanoscale},}\ }\href {\doibase 10.1016/j.physrep.2004.11.002} {\bibfield
  {journal} {\bibinfo  {journal} {Physics Reports}\ }\textbf {\bibinfo {volume}
  {406}},\ \bibinfo {pages} {379} (\bibinfo {year} {2005})}\BibitemShut
  {NoStop}%
\bibitem [{\citenamefont {Sambe}(1973)}]{Sambe1973}%
  \BibitemOpen
  \bibfield  {author} {\bibinfo {author} {\bibfnamefont {H.}~\bibnamefont
  {Sambe}},\ }\bibfield  {title} {\enquote {\bibinfo {title} {Steady states and
  quasienergies of a quantum-mechanical system in an oscillating field},}\
  }\href {\doibase 10.1103/PhysRevA.7.2203} {\bibfield  {journal} {\bibinfo
  {journal} {Phys. Rev. A}\ }\textbf {\bibinfo {volume} {7}},\ \bibinfo {pages}
  {2203} (\bibinfo {year} {1973})}\BibitemShut {NoStop}%
\bibitem [{\citenamefont {Shirley}(1965)}]{Shirley1965}%
  \BibitemOpen
  \bibfield  {author} {\bibinfo {author} {\bibfnamefont {J.~H.}\ \bibnamefont
  {Shirley}},\ }\bibfield  {title} {\enquote {\bibinfo {title} {Solution of the
  schr\"odinger equation with a hamiltonian periodic in time},}\ }\href
  {\doibase 10.1103/PhysRev.138.B979} {\bibfield  {journal} {\bibinfo
  {journal} {Phys. Rev.}\ }\textbf {\bibinfo {volume} {138}},\ \bibinfo {pages}
  {B979} (\bibinfo {year} {1965})}\BibitemShut {NoStop}%
\bibitem [{\citenamefont {Groth}\ \emph {et~al.}(2014)\citenamefont {Groth},
  \citenamefont {Wimmer}, \citenamefont {Akhmerov},\ and\ \citenamefont
  {Waintal}}]{Groth2014}%
  \BibitemOpen
  \bibfield  {author} {\bibinfo {author} {\bibfnamefont {C.~W.}\ \bibnamefont
  {Groth}}, \bibinfo {author} {\bibfnamefont {M.}~\bibnamefont {Wimmer}},
  \bibinfo {author} {\bibfnamefont {A.~R.}\ \bibnamefont {Akhmerov}}, \ and\
  \bibinfo {author} {\bibfnamefont {X.}~\bibnamefont {Waintal}},\ }\bibfield
  {title} {\enquote {\bibinfo {title} {Kwant: a software package for quantum
  transport},}\ }\href {http://stacks.iop.org/1367-2630/16/i=6/a=063065}
  {\bibfield  {journal} {\bibinfo  {journal} {New Journal of Physics}\ }\textbf
  {\bibinfo {volume} {16}},\ \bibinfo {pages} {063065} (\bibinfo {year}
  {2014})}\BibitemShut {NoStop}%
\bibitem [{\citenamefont {Foa~Torres}\ and\ \citenamefont
  {Roche}(2006)}]{FoaTorres2006}%
  \BibitemOpen
  \bibfield  {author} {\bibinfo {author} {\bibfnamefont {L.~E.~F.}\
  \bibnamefont {Foa~Torres}}\ and\ \bibinfo {author} {\bibfnamefont
  {S.}~\bibnamefont {Roche}},\ }\bibfield  {title} {\enquote {\bibinfo {title}
  {Inelastic quantum transport and peierls-like mechanism in carbon
  nanotubes},}\ }\href {http://link.aps.org/doi/10.1103/PhysRevLett.97.076804}
  {\bibfield  {journal} {\bibinfo  {journal} {Phys. Rev. Lett.}\ }\textbf
  {\bibinfo {volume} {97}},\ \bibinfo {pages} {076804} (\bibinfo {year}
  {2006})}\BibitemShut {NoStop}%
\bibitem [{\citenamefont {Foa~Torres}\ \emph {et~al.}(2008)\citenamefont
  {Foa~Torres}, \citenamefont {Avriller},\ and\ \citenamefont
  {Roche}}]{FoaTorres2008}%
  \BibitemOpen
  \bibfield  {author} {\bibinfo {author} {\bibfnamefont {L.~E.~F.}\
  \bibnamefont {Foa~Torres}}, \bibinfo {author} {\bibfnamefont
  {R.}~\bibnamefont {Avriller}}, \ and\ \bibinfo {author} {\bibfnamefont
  {S.}~\bibnamefont {Roche}},\ }\bibfield  {title} {\enquote {\bibinfo {title}
  {Nonequilibrium energy gaps in carbon nanotubes: Role of phonon
  symmetries},}\ }\href {http://link.aps.org/doi/10.1103/PhysRevB.78.035412}
  {\bibfield  {journal} {\bibinfo  {journal} {Phys. Rev. B}\ }\textbf {\bibinfo
  {volume} {78}},\ \bibinfo {pages} {035412} (\bibinfo {year}
  {2008})}\BibitemShut {NoStop}%
\bibitem [{\citenamefont {Calvo}\ \emph {et~al.}(2011)\citenamefont {Calvo},
  \citenamefont {Pastawski}, \citenamefont {Roche},\ and\ \citenamefont
  {Foa~Torres}}]{Calvo2011}%
  \BibitemOpen
  \bibfield  {author} {\bibinfo {author} {\bibfnamefont {H.~L.}\ \bibnamefont
  {Calvo}}, \bibinfo {author} {\bibfnamefont {H.~M.}\ \bibnamefont
  {Pastawski}}, \bibinfo {author} {\bibfnamefont {S.}~\bibnamefont {Roche}}, \
  and\ \bibinfo {author} {\bibfnamefont {L.~E.~F.}\ \bibnamefont
  {Foa~Torres}},\ }\bibfield  {title} {\enquote {\bibinfo {title} {Tuning
  laser-induced band gaps in graphene},}\ }\href
  {http://dx.doi.org/10.1063/1.3597412} {\bibfield  {journal} {\bibinfo
  {journal} {Appl. Phys. Lett.}\ }\textbf {\bibinfo {volume} {98}},\ \bibinfo
  {pages} {232103} (\bibinfo {year} {2011})}\BibitemShut {NoStop}%
\bibitem [{\citenamefont {G\'omez-Le\'on}\ \emph {et~al.}(2014)\citenamefont
  {G\'omez-Le\'on}, \citenamefont {Delplace},\ and\ \citenamefont
  {Platero}}]{Gomez-Leon2014}%
  \BibitemOpen
  \bibfield  {author} {\bibinfo {author} {\bibfnamefont {A.}~\bibnamefont
  {G\'omez-Le\'on}}, \bibinfo {author} {\bibfnamefont {P.}~\bibnamefont
  {Delplace}}, \ and\ \bibinfo {author} {\bibfnamefont {G.}~\bibnamefont
  {Platero}},\ }\bibfield  {title} {\enquote {\bibinfo {title} {Engineering
  anomalous quantum hall plateaus and antichiral states with ac fields},}\
  }\href {http://link.aps.org/doi/10.1103/PhysRevB.89.205408} {\bibfield
  {journal} {\bibinfo  {journal} {Phys. Rev. B}\ }\textbf {\bibinfo {volume}
  {89}},\ \bibinfo {pages} {205408} (\bibinfo {year} {2014})}\BibitemShut
  {NoStop}%
\bibitem [{\citenamefont {Perez-Piskunow}\ \emph {et~al.}(2015)\citenamefont
  {Perez-Piskunow}, \citenamefont {Foa~Torres},\ and\ \citenamefont
  {Usaj}}]{Perez-Piskunow2015}%
  \BibitemOpen
  \bibfield  {author} {\bibinfo {author} {\bibfnamefont {P.~M.}\ \bibnamefont
  {Perez-Piskunow}}, \bibinfo {author} {\bibfnamefont {L.~E.~F.}\ \bibnamefont
  {Foa~Torres}}, \ and\ \bibinfo {author} {\bibfnamefont {G.}~\bibnamefont
  {Usaj}},\ }\bibfield  {title} {\enquote {\bibinfo {title} {Hierarchy of
  floquet gaps and edge states for driven honeycomb lattices},}\ }\href
  {http://link.aps.org/doi/10.1103/PhysRevA.91.043625} {\bibfield  {journal}
  {\bibinfo  {journal} {Phys. Rev. A}\ }\textbf {\bibinfo {volume} {91}},\
  \bibinfo {pages} {043625} (\bibinfo {year} {2015})}\BibitemShut {NoStop}%
\bibitem [{\citenamefont {Delplace}\ \emph {et~al.}(2011)\citenamefont
  {Delplace}, \citenamefont {Ullmo},\ and\ \citenamefont
  {Montambaux}}]{Delplace2011}%
  \BibitemOpen
  \bibfield  {author} {\bibinfo {author} {\bibfnamefont {P.}~\bibnamefont
  {Delplace}}, \bibinfo {author} {\bibfnamefont {D.}~\bibnamefont {Ullmo}}, \
  and\ \bibinfo {author} {\bibfnamefont {G.}~\bibnamefont {Montambaux}},\
  }\bibfield  {title} {\enquote {\bibinfo {title} {Zak phase and the existence
  of edge states in graphene},}\ }\href
  {http://link.aps.org/doi/10.1103/PhysRevB.84.195452} {\bibfield  {journal}
  {\bibinfo  {journal} {Phys. Rev. B}\ }\textbf {\bibinfo {volume} {84}},\
  \bibinfo {pages} {195452} (\bibinfo {year} {2011})}\BibitemShut {NoStop}%
\bibitem [{\citenamefont {Xiao}\ \emph {et~al.}(2015)\citenamefont {Xiao},
  \citenamefont {Ma}, \citenamefont {Yang}, \citenamefont {Sheng},
  \citenamefont {Zhang},\ and\ \citenamefont {Chan}}]{Xiao2015}%
  \BibitemOpen
  \bibfield  {author} {\bibinfo {author} {\bibfnamefont {M.}~\bibnamefont
  {Xiao}}, \bibinfo {author} {\bibfnamefont {G.}~\bibnamefont {Ma}}, \bibinfo
  {author} {\bibfnamefont {Z.}~\bibnamefont {Yang}}, \bibinfo {author}
  {\bibfnamefont {P.}~\bibnamefont {Sheng}}, \bibinfo {author} {\bibfnamefont
  {Z.~Q.}\ \bibnamefont {Zhang}}, \ and\ \bibinfo {author} {\bibfnamefont
  {C.~T.}\ \bibnamefont {Chan}},\ }\bibfield  {title} {\enquote {\bibinfo
  {title} {Geometric phase and band inversion in periodic acoustic systems},}\
  }\href {http://dx.doi.org/10.1038/nphys3228} {\bibfield  {journal} {\bibinfo
  {journal} {Nat. Phys.}\ }\textbf {\bibinfo {volume} {11}},\ \bibinfo {pages}
  {240} (\bibinfo {year} {2015})}\BibitemShut {NoStop}%
\bibitem [{\citenamefont {Sherson}\ \emph {et~al.}(2010)\citenamefont
  {Sherson}, \citenamefont {Weitenberg}, \citenamefont {Endres}, \citenamefont
  {Cheneau}, \citenamefont {Bloch},\ and\ \citenamefont {Kuhr}}]{Sherson2010}%
  \BibitemOpen
  \bibfield  {author} {\bibinfo {author} {\bibfnamefont {J.~F.}\ \bibnamefont
  {Sherson}}, \bibinfo {author} {\bibfnamefont {C.}~\bibnamefont {Weitenberg}},
  \bibinfo {author} {\bibfnamefont {M.}~\bibnamefont {Endres}}, \bibinfo
  {author} {\bibfnamefont {M.}~\bibnamefont {Cheneau}}, \bibinfo {author}
  {\bibfnamefont {I.}~\bibnamefont {Bloch}}, \ and\ \bibinfo {author}
  {\bibfnamefont {S.}~\bibnamefont {Kuhr}},\ }\bibfield  {title} {\enquote
  {\bibinfo {title} {Single-atom-resolved fluorescence imaging of an atomic
  mott insulator},}\ }\href {\doibase http://dx.doi.org/10.1038/nature09378}
  {\bibfield  {journal} {\bibinfo  {journal} {Nature}\ }\textbf {\bibinfo
  {volume} {467}},\ \bibinfo {pages} {68} (\bibinfo {year} {2010})}\BibitemShut
  {NoStop}%
\bibitem [{\citenamefont {Bakr}\ \emph {et~al.}(2010)\citenamefont {Bakr},
  \citenamefont {Peng}, \citenamefont {Tai}, \citenamefont {Ma}, \citenamefont
  {Simon}, \citenamefont {Gillen}, \citenamefont {Fölling}, \citenamefont
  {Pollet},\ and\ \citenamefont {Greiner}}]{Bakr2010}%
  \BibitemOpen
  \bibfield  {author} {\bibinfo {author} {\bibfnamefont {W.~S.}\ \bibnamefont
  {Bakr}}, \bibinfo {author} {\bibfnamefont {A.}~\bibnamefont {Peng}}, \bibinfo
  {author} {\bibfnamefont {M.~E.}\ \bibnamefont {Tai}}, \bibinfo {author}
  {\bibfnamefont {R.}~\bibnamefont {Ma}}, \bibinfo {author} {\bibfnamefont
  {J.}~\bibnamefont {Simon}}, \bibinfo {author} {\bibfnamefont {J.~I.}\
  \bibnamefont {Gillen}}, \bibinfo {author} {\bibfnamefont {S.}~\bibnamefont
  {Fölling}}, \bibinfo {author} {\bibfnamefont {L.}~\bibnamefont {Pollet}}, \
  and\ \bibinfo {author} {\bibfnamefont {M.}~\bibnamefont {Greiner}},\
  }\bibfield  {title} {\enquote {\bibinfo {title} {Probing the
  superfluid–to–mott insulator transition at the single-atom level},}\
  }\href {\doibase http://dx.doi.org/10.1126/science.1192368} {\bibfield
  {journal} {\bibinfo  {journal} {Science}\ }\textbf {\bibinfo {volume}
  {329}},\ \bibinfo {pages} {547} (\bibinfo {year} {2010})}\BibitemShut
  {NoStop}%
\end{thebibliography}

%

\end{document}